\documentclass[preprint, showpacs, reprintnumbers, amsmath, amssymb, superscriptaddress]{revtex4-2}

\usepackage{epsf}
\usepackage{graphicx}
\usepackage{sidecap}

 \usepackage{soul}
 \usepackage{array}
 \usepackage{amsmath}
 \usepackage{amssymb}
 \usepackage{dcolumn}
 \usepackage{epstopdf}
 \usepackage{bm}
 \usepackage{gensymb}
 \usepackage{amsmath}

\usepackage{color}
\usepackage{hyperref}
\usepackage{soul}
\sethlcolor{green}
\hypersetup{
    colorlinks=true,
    citecolor=red,
    linkcolor=red,
    filecolor=blue,   
    urlcolor=blue,
}
 
\def \beq {\begin{equation}}
\def \eeq {\end{equation}}
\begin{document}

\onecolumngrid

\begin{center}
 
  \textbf{\Large Electronic structure in a rare-earth based nodal-line semimetal candidate PrSbTe}\\[.2cm]
Sabin~Regmi$^{1,2,*}$, Iftakhar~Bin~Elius$^{1,*}$, Anup~Pradhan~Sakhya$^{1}$, Milo~Sprague$^{1}$, Mazharul~Islam~Mondal$^{1}$,  Nathan~Valadez$^{1}$, Volodymyr~Buturlim$^3$, Kali~Booth$^1$, Tetiana Romanova$^4$, Krzysztof~Gofryk$^2$, Andrzej~Ptok$^{5}$, Dariusz~Kaczorowski$^{4}$, Madhab~Neupane$^{1,{\dagger}}$\\[.2cm]
 {\itshape
    $^{1}$Department of Physics, University of Central Florida, Orlando, Florida  32816, USA\\
    $^{2}$Center for Quantum Actinide Science and Technology,\\ Idaho National Laboratory, Idaho Falls, Idaho 83415, USA\\
    $^{3}$Glenn T. Seaborg Institute, Idaho National Laboratory, Idaho Falls, Idaho 83415, USA\\
    $^{4}$Institute of Low Temperature and Structure Research, \\Polish Academy of Sciences, ul. Ok\'{o}lna 2, 50-422 Wroc\l{}aw, Poland\\
  	$^{5}$Institute of Nuclear Physics, Polish Academy of Sciences, \\W. E. Radzikowskiego 152, PL-31342 Krak\'{o}w, Poland
  }
\\[.2cm]
$^{*}$These authors contributed equally to this work.\\
$^{\dagger}$Corresponding author: madhab.neupane@ucf.edu
\\[1cm]
\end{center}

\begin{abstract}
\textbf{Nodal-line semimetals feature topologically protected band crossings between the bulk valence and conduction bands that extend along a finite dimension in the form of a line or a loop. While ZrSiS and similar materials have attracted extensive research as hosts for the nodal-line semimetallic phase, an alternative avenue has emerged in the form of isostructural rare-earth ($RE$)-based $RE$SbTe  materials. Such systems possess intriguing potentialities for harboring elements of magnetic ordering and electronic correlations owing to the presence of 4$f$ electrons intrinsic to the $RE$ elements. In this study, we have carried out angle-resolved photoemission spectroscopy (ARPES) and thermodynamic measurements in conjunction with first-principles computations on PrSbTe to elucidate its electronic structure and topological characteristics. Magnetic and thermal  characterizations indicate the presence of well-localized $4f$ states with the absence of any discernible phase transition down to 2~K. The ARPES results reveal the presence of gapless Dirac crossings that correspond to a nodal-line along the $X-R$ direction in the three-dimensional Brillouin zone. Furthermore, Dirac crossing that makes up nodal line, which forms a diamond-shaped nodal plane centered at the center of the Brillouin zone is also identified within the experimental resolution. This study on the electronic structure of  PrSbTe contributes to the understanding of the pivotal role played by spin-orbit coupling in the context of the $RE$SbTe family of materials.}
\end{abstract}
\maketitle

The prediction and subsequent discovery of topologically protected surface states in three-dimensional (3D) topological insulator (TI) \cite{Hasan2010, Qi2011}  not only instigated the search for such materials but also led to the realization of topology in a diverse range of quantum materials beyond TIs, including topological semimetals \cite{Armitage2018} and superconductors \cite{Qi2011, Sato2017}. In topological semimetals,  the protected band crossing occurs between the bulk valence and conduction bands themselves \cite{Armitage2018}, unlike the protected band crossings between spin-polarized linear surface states that bridge the gapped out bulk valence and conduction bands in TIs. Such topological protection is supported by the concurrent presence of time-reversal (TR) and inversion (I) symmetries in what are referred to as Dirac semimetals with a distinctive four-fold degeneracy at the band crossing because of the Kramer's degeneracy principle \cite{Liu2014, Neupane2014}. On the other hand, Weyl semimetals manifest a pair of two-fold band crossings with opposite topological charge or chirality \cite{Xu2015, Lv2015, Soluyanov2015, Yan2017}. The presence of extra crystalline symmetries may lead to a distinct scenario where the protected band crossings extend along one or two dimensions, leading to the emergence of a topological nodal-line semimetallic phase \cite{Burkov2011, Bian2016, Neupane2016, Schoop2016, Yang2018}.\\
\begin{figure*}
\includegraphics[width=1\textwidth]{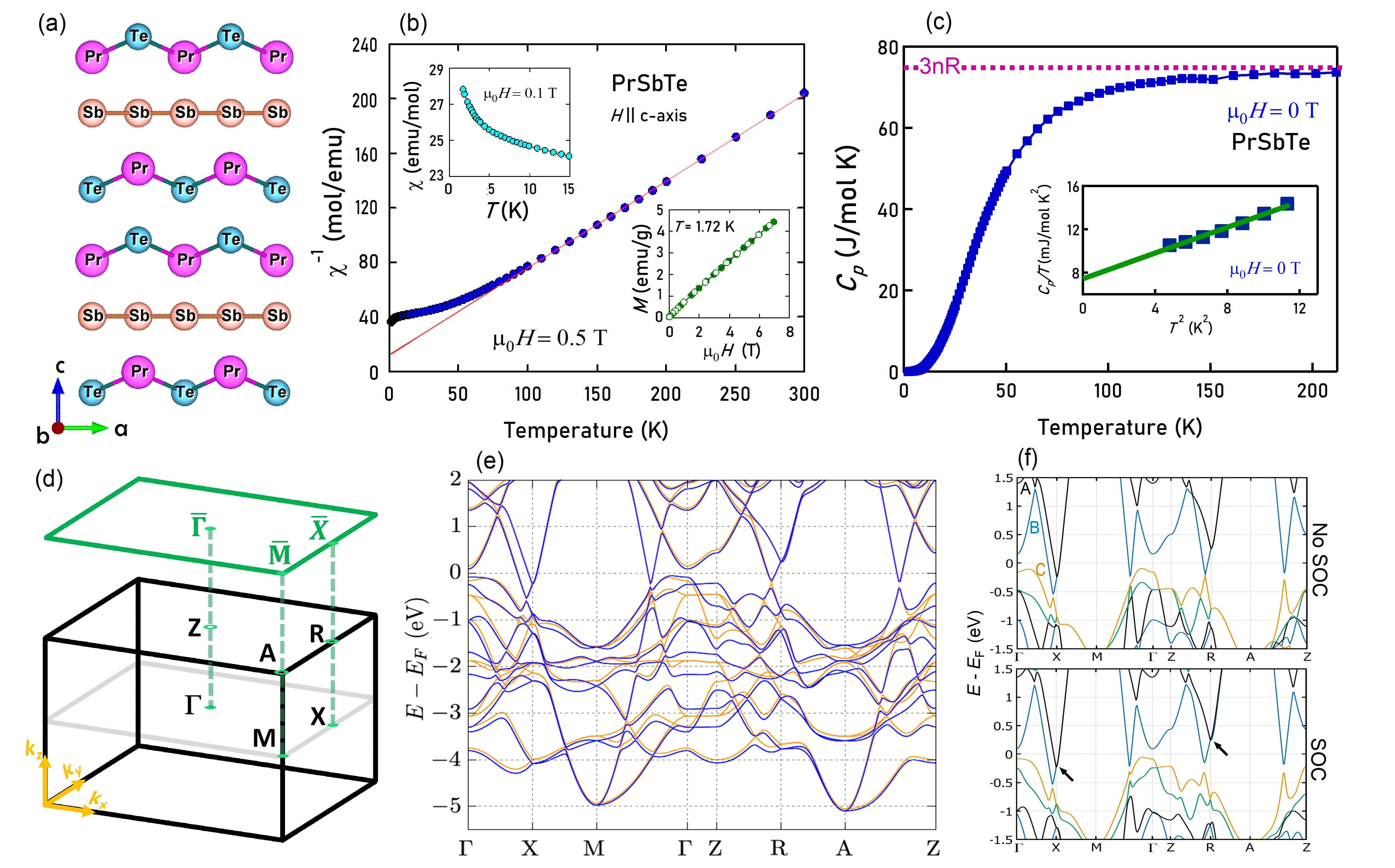}
\caption{Crystal structure, bulk characterization, and bulk band structure. (a) Side view of the crystal structure. (b) Inverse magnetic susceptibility as a function of temperature. Inset on top left shows the low-temperature magnetic susceptibility data measured in a magnetic field of 0.1~T. Inset on bottom right shows magnetization as a function of applied field taken at 1.72 K, with increasing (full circles) and decreasing (open circles) field. (c) Temperature variation of heat capacity measured in zero field. Inset: Low temperature plot  of ratio of heat capacity to temperature versus temperature squared and its linear fit. (d) Bulk and (001) surface  Brillouin zones. (e) Calculated band structure without (orange) and with (blue) SOC along various high-symmetry directions. (f) Band structure focused on bands A, B, and C when SOC is not taken into account (top) and  is taken into account (bottom).}
\label{F1}
\end{figure*}
A well-known symmetry that is known to preserve the topological crossing in nodal-line semimetals is the nonsymmorphic symmetry, which is evident in the celebrated nodal-line semimetal ZrSiS and analogous $M$Si$C$ ($M$ = Zr/Hf; $C$ = S/Se/Te) materials with a Si square lattice configuration in the crystal structure \cite{Neupane2016, Schoop2016, Yang2018, Topp2016, Hu2016, Takane2016, Hosen2017, Chen2017, Lou2016, Hosen2018, Fu2019}. These materials have relatively modest spin-orbit coupling (SOC) and are non-magnetic. To delve into the response of nodal-line physics to the varying degrees of the SOC as well as to the introduction of magnetism, it is necessary to explore materials containing heavier elements, particularly  those which can introduce long-range magnetic order.  Recently, a promising avenue has emerged in a family of isostructural rare-earth ($RE$)-based $RE$SbTe  materials, in which the presence of $RE$ elements introduces 4$f$ electrons offering an opportunity to investigate the interplay between strong electronic interactions, magnetically ordered moments, and topological properties. The presence of magnetically ordered phase has been reported in several members of the $RE$SbTe family  \cite{Chen2017Ce, Hosen2018Gd, Schoop2018,  Sankar2019, Yang2020, Lv2019, Pandey2020,  Pandey2021, Plokhikh2022, Gao2022, Regmi2023, Gao2023}. Additionally, signatures of Kondo localization have been observed in various members of this family, indicating the presence of strong electronic correlation effects \cite{Chen2017Ce, Lv2019, Li2021, Pandey2020, Pandey2021}. Studies of the electronic band structure have revealed that the choice of the $RE$ element enables  tuning of both topology and overall band structure. Materials with lighter $RE$ elements like (La/Nd/Sm)SbTe  \cite{Pandey2021, Wang2021, Regmi2022, Regmi2023} exhibit gapless topological band-crossings and the heavier counterparts such as (Ho/Dy)SbTe \cite{Yue2020, Yang2020, Gao2022, Shumiya2022} possess strong SOC-induced gap in their electronic structure, which could potentially lead to the emergence of a weak topological insulator state. A Dirac state is protected within the TR symmetry-broken magnetic phase of GdSbTe, with the protection facilitated by the combination of broken TR and crystalline rotoinversion symmetries \cite{Hosen2018Gd}. Furthermore, with tunability of the 4$f$ moments, CeSbTe can accommodate a variety of topological phases \cite{Schoop2018, Topp2019}. These collective findings underscore the role SOC plays on the electronic structure and topological characteristics within the $RE$SbTe family and the significance of extending research studies to encompass additional members of this family. These studies will help comprehend the role SOC plays in shaping the electronic and topological properties within this material family. \\

\begin{figure*}
\includegraphics[width=1\textwidth]{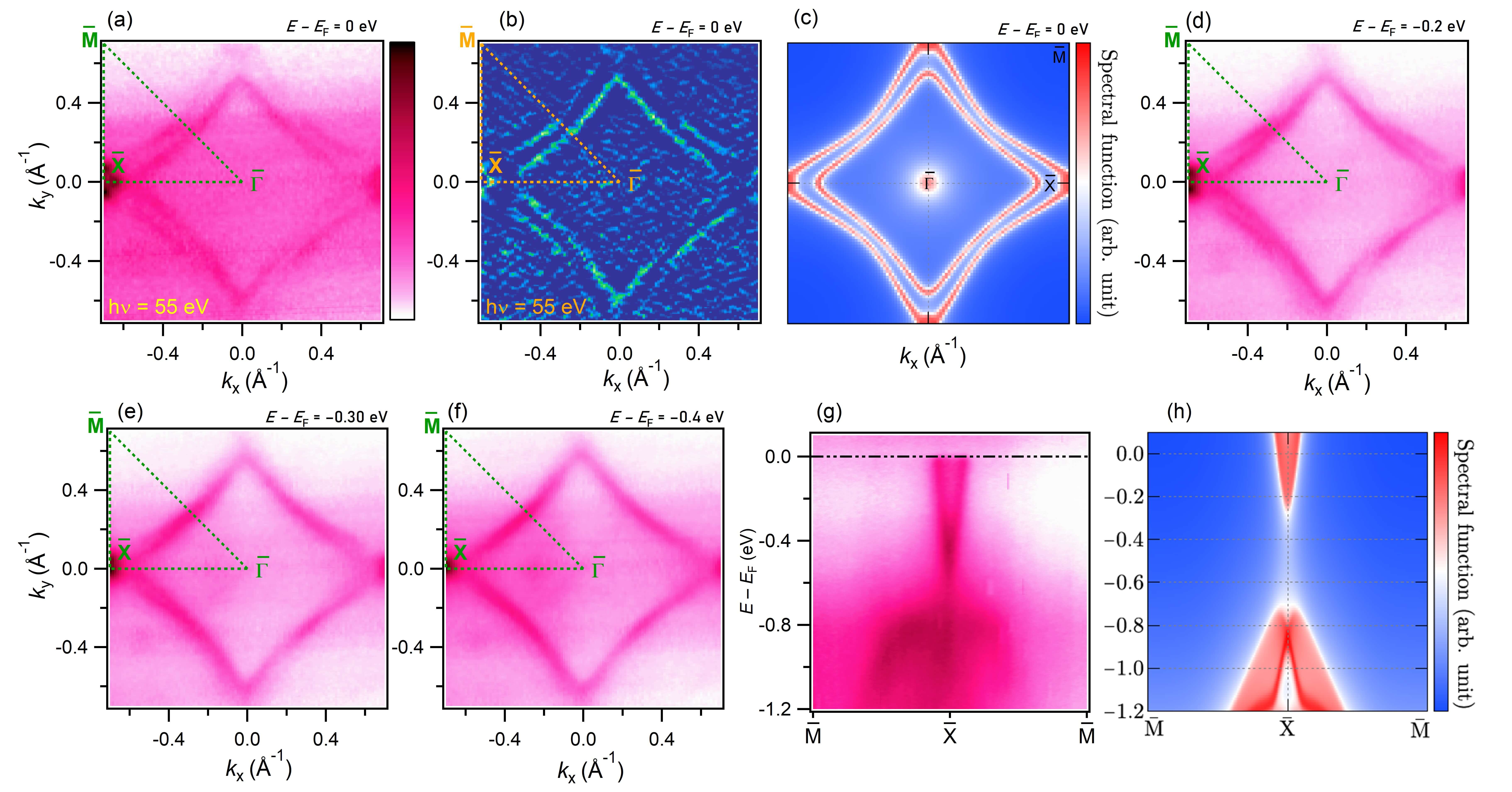}
\caption{Energy contours and $\overline{M}-\overline{X}-\overline{M}$ band structure. (a) Fermi surface measured with a photon energy of 55 eV. (b) Second derivative plot of (a). (c) Calculated FS spectrum on the (001) surface. (d)-(f) Constant energy contours plots at mentioned binding energy values. (g) ARPES band structure along the $\overline{M}-\overline{X}-\overline{M}$ direction. (h) Calculated (001) surface spectrum along $\overline{M}-\overline{X}-\overline{M}$. Experimental data were collected at the ALS 10.0.1.1 beamline at a temperature of 15 K.}
\label{F2}
\end{figure*}
Despite the growing interest directed towards $RE$SbTe materials, as evident from recent investigations, PrSbTe within this family remains relatively unexplored. In the present study, we have carried out angle-resolved photoemission spectroscopy (ARPES) measurements on PrSbTe with an objective of elucidating its as-yet unveiled electronic structure. These measurements have been complemented by thermodynamic studies and first-principles calculations, performed to establish or explain the experimental findings. The ARPES results reveal Dirac-like crossings along the $\overline{\Gamma}-\overline{M}$ and $\overline{\Gamma}-\overline{X}$ directions, which are gapless within the experimental resolution. These Dirac-like crossings are attributed to two distinct nodal-lines, one along the $X-R$ direction and the other transversing the $\Gamma-M$ direction. The later nodal-line delineates a diamond shaped nodal plane centered at the $\Gamma$ point.   This study brings out the electronic band structure of an additional material candidate in the $RE$SbTe family and facilitates the understanding of the role of SOC and $RE$ elemental choice in shaping the  topological electronic structure within this family of materials.\\

\noindent \textit{Methods-}~ PrSbTe single crystals of high-quality were synthesized using chemical vapor transport technique and characterized for chemical composition and crystal structure using energy dispersive X-ray spectroscopy (EDX) and  X-ray diffraction (XRD), respectively. Magnetic and heat capacity measurements were performed using Quantum Design MPMS-XL SQUID magnetometer and Quantum Design PPMS-14 system, respectively. ARPES experiments were conducted at the Advanced Light Source (ALS) beamline 10.0.1.1 at a temperature of 15~K and at the Stanford Synchrotron Radiation Lightsource (SSRL) beamline 5-2 at a temperature 7~K. The energy resolution was maintained below 20~meV for all ARPES measurements.  First-principles calculations were performed within the framework of  density functional theory (DFT) \cite{Hohenberg1964, Kohn1965}, employing simultaneously within the Vienna \textit{ab initio} simulation package \cite{Kresse1994, Kresse1996, Kresse1999} and Quantum Espresso \cite{Giannozzi2009, Giannozzi2017, Giannozzi2020} on the basis of projector augmented wave potential \cite{Blochl1994}. For more details on experimental and computational methodologies, see sections S1 and S2 in the Supplemental Material (SM) accompanying this manuscript \cite{SM}.\\

\begin{figure*}
\includegraphics[width=1\textwidth]{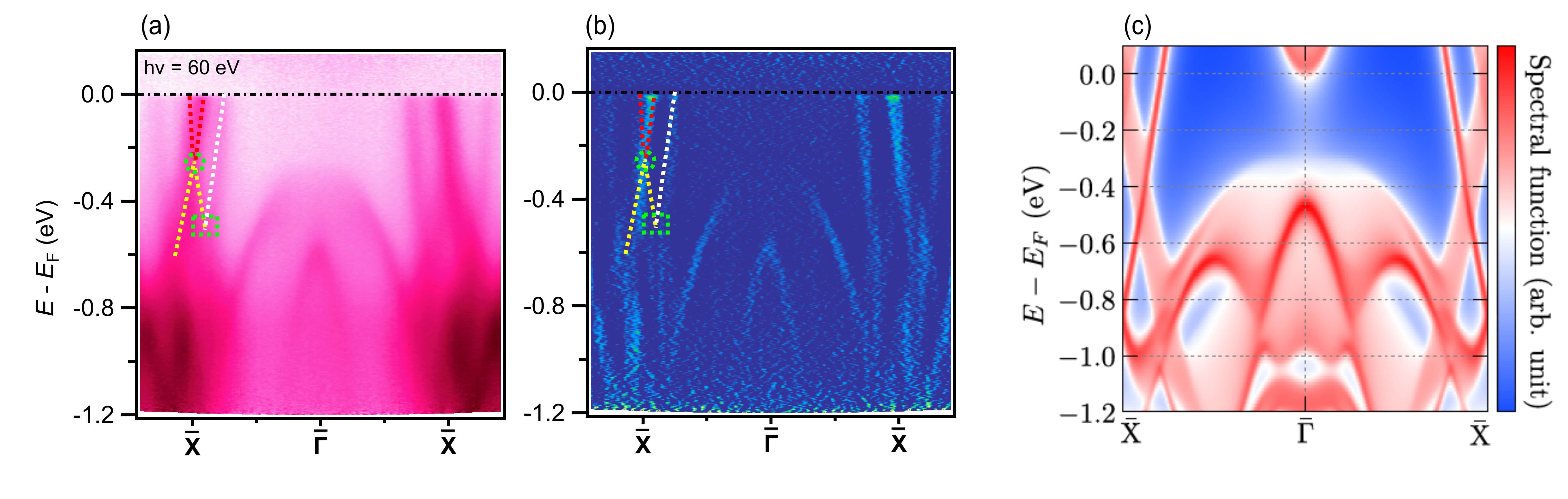}
\caption{Band structure along the $\overline{X}-\overline{\Gamma}-\overline{X}$ direction. (a) $\overline{X}-\overline{\Gamma}-\overline{X}$ electronic structure map measured with 60 eV linear horizontal polarized light. (b) Second derivative plot of (a). (c) Calculated surface spectrum along $\overline{X}-\overline{\Gamma}-\overline{X}$ on the (001) surface. Experimental data were collected at the SSRL 5-2 beamline at a temperature of 7 K using linear horizontal polarized light.}
\label{F3}
\end{figure*}

\noindent \textit{Results and discussion-}~PrSbTe, similar to other materials belonging to the $RE$SbTe family, adopts a PbFCl-type crystal structure within the $P4/nmm$ space group. A graphical representation of the side view of the crystal structure of PrSbTe is presented in Fig. \ref{F1}(a). The layers of Sb atoms are arranged in square lattice configuration and sandwiched by the Pr-Te layers. Figure~\ref{F1}(b) outlines the magnetic properties of single-crystalline PrSbTe, where the magnetic field is aligned parallel to the tetragonal axis. Above about 100 K, the compound manifests Curie-Weiss behavior, exhibiting effective magnetic moment  ($\mu_{eff}$) of 3.54(3)~$\mu_B$ and a paramagnetic Curie temperature ($\theta_p$) of -18(1)~K. At lower temperatures, the inverse magnetic susceptibility ($\chi^{-1}(T)$) departs significantly from a linear trend, likely due to crystalline electric field interactions. The experimental value of $\mu_{eff}$ closely approximates the theoretical value (3.58~$\mu_B$) predicted within the Russell-Saunders coupling scenario for a free Pr$^{3+}$ ion. The notably negative value of $\theta_p$ indicates strong antiferromagnetic exchange interactions. Nevertheless, as can be inferred from the upper inset to Fig.~\ref{F1}(b), PrSbTe remains in a paramagnetic state down to a temperature of 1.72~K, similar to previously reported \cite{Plokhikh2023}. However, an upturn in $\chi (T)$ seen at the lowest temperatures covered hints at a possible magnetic phase transition below the limit in the thermodynamic measurements performed in this study. The paramagnetic character of the compound at $T$ = 1.72~K is further corroborated by a quasi-linear dependence of the magnetization without any sign of hysteresis or metamagnetic transition (see the lower inset to Fig.~\ref{F1}(b)). The temperature variation of the heat capacity ($C_p$) (Fig.~\ref{F1}(c)) does not show any anomalies, in line with the absence of magnetic phase transition in the magnetic measurements. At high temperatures, the heat capacity approaches the theoretical Dulong-Petit limit of $3nR$, where $n$ = 3 (number of atoms in the  formula unit) and $R$ = 8.314~mJ/(mol~K). Applying a large magnetic field of up to 10~T does not bring any significant alterations in the $C_p(T)$ curve (See Fig.~S2 in the SM \cite{SM}). At low temperatures, the temperature variation of the heat capacity adheres to the equation $C_p(T) = \gamma T + \beta T^3$. By employing linear fitting to the plot of ${C_p}/{T}$ versus $T^2$ (see inset), the derived parameters are $\gamma$ = 7.45~mJ/(mol K$^2$), indicative of modest electronic contribution to the heat capacity, and $\beta$ = 0.59~mJ/(mol K$^4$). By using the $\beta$ value, the Debye temperature is estimated to be $\Theta_D = \left[ {12\pi^4nR}/{(5\beta)}\right]^{1/3}~ \sim$ 215~K, which is in the range of the values reported for other $RE$SbTe compounds, such as 195~K in GdSbTe \cite{Sankar2019}, 207~K in TbSbTe \cite{Gao2023}, 216~K in DySbTe \cite{Gao2022},  236.6~K in NdSbTe \cite{Pandey2020}, 244.3~K in LaSbTe \cite{Pandey2020}, 232.14~K in SmSbTe \cite{Pandey2021}.\\

Figure~\ref{F1}(d) presents an illustration of the three-dimensional bulk Brillouin zone (BZ) and its projection onto the (001) surface, with the high-symmetry points on both BZs marked. The bulk band structures along various high-symmetry directions are depicted in Fig. \ref{F1}(e). The bands are differentiated by color, to allow identification and comparison of the band structures before (orange) and after (blue) including the effect of the spin-orbit coupling (SOC). Notably, the consideration of SOC influences a notable change in the overall electronic structure. Here, we focus on three bands labeled A, B, and C (see Fig.~\ref{F1}(f)). In the absence of SOC, bands B and C form a gapped state with a modest gap along the $\Gamma-X$ direction, however, a nearly gapless state along the $Z-R$ direction. Similarly, bands A and B manifest gapless crossings at the $X$ and $R$ points, with the former lying  below the Fermi level and the latter above it. Upon the introduction of SOC, the gapped state formed by bands B and C along the $\Gamma-X$ direction remains relatively unaltered. However, a pronounced gap forms at the $k_z=\pi$ equivalent state (i.e., along $Z-R$). Shifting the focus to the crossings between bands A and B, one can observe that the inclusion of SOC does not disrupt the inherent gapless nature of the crossings at the $X$ and $R$ points. These crossings form a gapless nodal-line configuration extending along the bulk $X-R$ direction, commonly reported in some lighter members of the $RE$SbTe family \cite{Regmi2023, Wang2021, Regmi2022}. Along the $\Gamma-M$ direction, a strong gap occurs between the bands B and C in the absence of SOC, which is significantly reduced upon the inclusion of SOC.\\

\begin{figure*}
\includegraphics[width=1\textwidth]{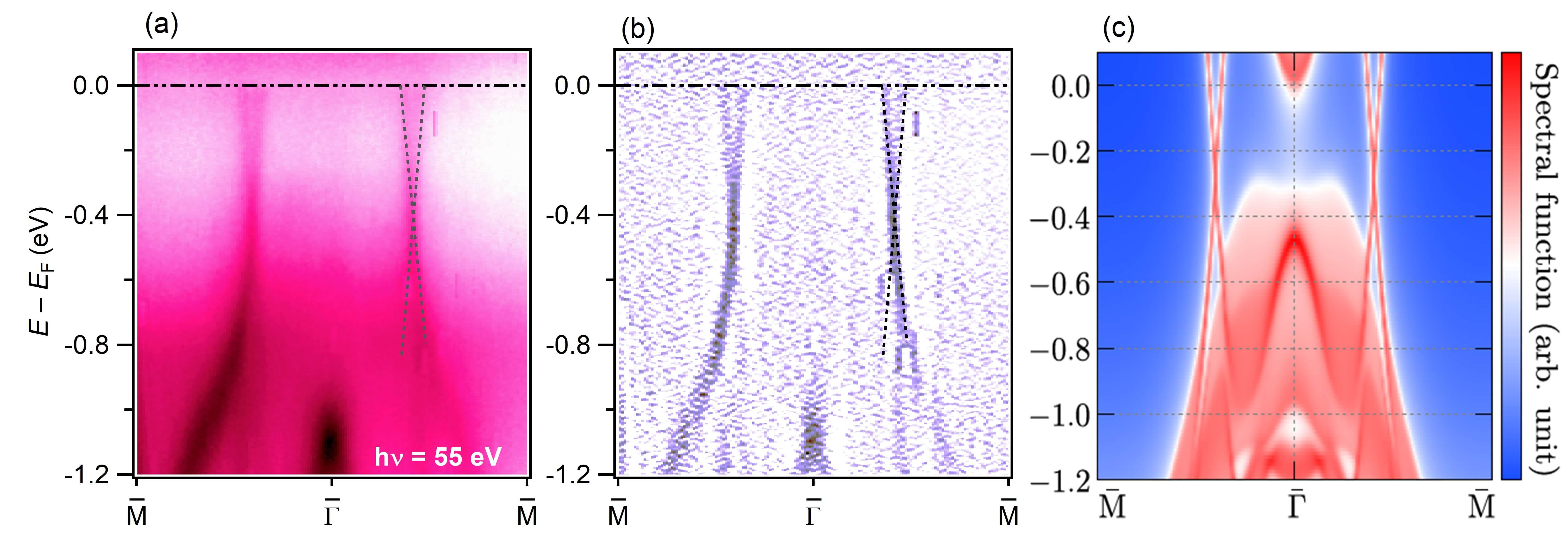}
\caption{Band structure along the $\overline{M}-\overline{\Gamma}-\overline{M}$ direction. (a) $\overline{M}-\overline{\Gamma}-\overline{M}$ electronic structure map measured with 55 eV photon energy. (b) Second derivative plot of (a). (c)  Calculated surface spectrum along $\overline{M}-\overline{\Gamma}-\overline{M}$ on the (001) surface. Experimental data were collected at the ALS 10.0.1.1 beamline at a temperature of 15 K.}
\label{F4}
\end{figure*}

Similar to ZrSiS-type tetragonal materials, the Fermi surface (FS) of PrSbTe also consists of diamond shaped pocket that is centered at the $\Gamma$ point (Fig. \ref{F2}(a)). This diamond pocket exhibits a double-sheet nature, clearly discernible from the experimental FS as well as its second derivative plot (Fig. \ref{F2}(b)). The FS spectrum obtained from DFT calculation resembles the experimental one quite well. Upon examining the constant energy contours at higher binding energies below the Fermi level, the separation of the sheets within the diamond shape reduces and eventually consolidates into a single sheet at a binding energy of around 400 meV (Figs. \ref{F2}(d-f)) [Also see Fig.~S5 in the SM \cite{SM}]. Additionally, a small pocket exists around the $\overline{X}$ point, which gives rise to a narrow electron-like band, which can be observed in the dispersion map along the $\overline{M}-\overline{X}-\overline{M}$ direction. By comparing with the theoretical calculation, we can establish that the electron-like band is a surface band, which is found in other $RE$SbTe and ZrSiS-type materials as well, and that it co-exists with other bulk bands. A similar band structure can be observed in the calculated surface spectrum along $\overline{M}-\overline{X}-\overline{M}$, where a significant gap appears at the  $\overline{X}$ point.\\

To gain deeper insight into the nodal-line topology in PrSbTe, we took the dispersion maps along the $\overline{X}-\overline{\Gamma}-\overline{X}$ and $\overline{M}-\overline{\Gamma}-\overline{M}$ directions and the results are presented in Figs.~\ref{F3} and \ref{F4}, respectively. A dispersion map taken with a photon energy of 60 eV along the $\overline{X}-\overline{\Gamma}-\overline{X}$ direction is illustrated in Fig. \ref{F3}(a) and its second derivative plot is shown in Fig. \ref{F3}(b). Below the Fermi level, two band-crossing features can be observed, one of which lies exactly at the $\overline{X}$ point (enclosed by a dashed circle)  and the other lies away from the $\overline{X}$ point towards the $\overline{\Gamma}$ point (enclosed by a dashed  square). The crossing at the $\overline{X}$ point is consistent with the anticipated gapless crossing between bands A and B, as predicted by the calculated band structure presented in Fig. \ref{F1}. This correspondence means that it is associated with the nodal-line that forms along the $X-R$ direction, which is the out-of-plane direction in our measurement setup. The location of this crossing is around 250 meV below the Fermi level at the photon energy of 60 eV [also see Fig.~S6 in the SM \cite{SM}]. The theoretical surface spectrum along the  $\overline{X}-\overline{\Gamma}-\overline{X}$, which is shown in Fig.~\ref{F3}(c), reasonably matches the experimental dispersion map. A nodal-line is formed by the bulk bands, which exhibits dispersion variations with changing photon energy.  The bulk nature of the bands involved in this nodal crossing is evidenced by taking  dispersion maps at different photon energies. A gapless crossing exists at each photon energies as well, further supporting the notion of a gapless nodal-line extending along the $k_z$ direction and the choice of different photon energy seems to change the dispersion of the bands and also varies the position of the crossing (see Figs.~S6 and S7 in the SM \cite{SM}).\\

The other crossing along the $\overline{\Gamma}-\overline{X}$ direction enclosed by a dashed square (around 540 meV below the Fermi level; see Fig.~\ref{F3}(a)) corresponds to the crossing made by bands B and C in Fig.~\ref{F1}(e,f). It is important to note that the bands B and C are responsible for the diamond shaped pocket in the Fermi surface and energy contours. In Fig.~\ref{F4}, we present the surface spectrum along the $\overline{M}-\overline{\Gamma}-\overline{M}$ direction, that intersects the diamond-shaped pocket. Figures~\ref{F4}(a-b) show the experimental spectrum and its second derivative plot, respectively. Two linear bands extending over a large energy range appear to cross one other at a binding energy of $\sim$ 400 meV. This is in concert with the energy contours presented in Fig.~\ref{F2}, where the two-sheets within the diamond-shaped pocket merge around this binding energy in this direction.  In the bulk band calculations (Fig.~\ref{F1}(e,f)), while SOC diminishes the gap between the bands B and C  along the $\Gamma-M$ direction, a small gap still exists, which remains unresolved in the experimental data. In Fig.~\ref{F4}(c), a calculated surface spectrum is presented, which fairly reproduces the experimental observation, including the gapless crossing. Considering the fact that the theoretical surface spectrum is produced by the projection from all $k_z$ planes onto the (001) surface, this slight discrepancy between the bulk band calculation and experiment may be attributed to the limitations of the experimental  $k_z$ resolution, further compounded by slight deviation in the chemical composition wherein Sb is partially replaced by Te (see Fig.~S1 in the SM \cite{SM}). Similar effect of partial replacement of Sb atoms by Te atoms has been recognized in other $RE$SbTe compounds as well \cite{Lv2019, Lei2019, Pandey2020} and could also be the reason behind the observed Fermi energy shift (of about 100 meV) between theoretical calculations and experimental observations. An additional factor could be the finite energy resolution intrinsic to the experimental setup. It is also essential to mention the potential of the discrepancy arising from the inability of DFT to precisely replicate the  experimental results, especially in metallic/semimetallic systems.\\ 

\noindent\textit{Conclusion-}~In conclusion, we performed an electronic structure study of PrSbTe through the combined utilization of ARPES and complementary DFT calculations. Multiple Dirac-like crossings have been identified in this material, which remain gapless within the experimental resolution. The Dirac crossing observed at the corner of the surface BZ corresponds to the projection of the nodal-line crossing along the bulk $X-R$ direction, a feature that remains gapless in both theoretical calculations and experimental observations. The Dirac-like crossing observed along the $\Gamma-M$ direction corresponds to nodal line that shapes the diamond pocket centered at $\Gamma$. While bulk band calculations predict a gapped nature for this crossing, experimental results reveal a gapless nature, within the limitations of the experimental resolution. This study provides an important platform for examining and comprehending the role of the choice of $RE$ and the degree of SOC in the electronic structure and topological characteristics within the family of $RE$SbTe systems. \\

\noindent\textit{Acknowledgment-}~M.N. acknowledges the support from the National Science Foundation under CAREER award DMR-1847962 and the Air Force Office of Scientific Research MURI Grant No. FA9550-20-1-0322. K.G. acknowledges support from the Division of Materials Science and Engineering, Office of Basic Energy Sciences, Office of Science of the U. S. Department of Energy (U.S. DOE).  S.R. and V.B. acknowledge the support from Idaho National Laboratory's Laboratory Directed Research and Development (LDRD) program under DOE Idaho Operations Office Contract DE-AC07-05ID14517.  D.K. and T.R. were supported by the National Science Centre (Poland) under research grant 2021/41/B/ST3/01141. A.P. acknowledges the support by National Science Centre (NCN, Poland) under Projects No. 2021/43/B/ST3/02166. This research used resources of the Advanced Light Source (ALS) at the Lawrence Berkeley National Laboratory, which is a DOE Office of Science User Facility under Contract No. DE-AC02-05CH11231. Use of the Stanford Synchrotron Radiation Lightsource (SSRL), SLAC National Accelerator Laboratory, is supported by the U.S. Department of Energy, Office of Science, Office of Basic Energy Sciences under Contract No. DE-AC02-76SF00515. We are grateful to Dr. Sung-Kwan Mo for the beamline assistance at ALS, and Dr. Makoto Hashimoto and Dr. Donghui Lu for beamline assistance at SSRL.

\clearpage

\setcounter{figure}{0}
\renewcommand{\figurename}{\textbf{Fig. S}}
\renewcommand{\thefigure}{{\textbf{\arabic{figure}}}}
\begin{center}
\textbf{\Large SUPPLEMENTAL MATERIAL}
\end{center}
 \section*{S1. Experimental techniques}
\subsubsection*{Crystal Structure and Sample Characterization}
\begin{figure*} [h!]
\includegraphics[width=1\textwidth]{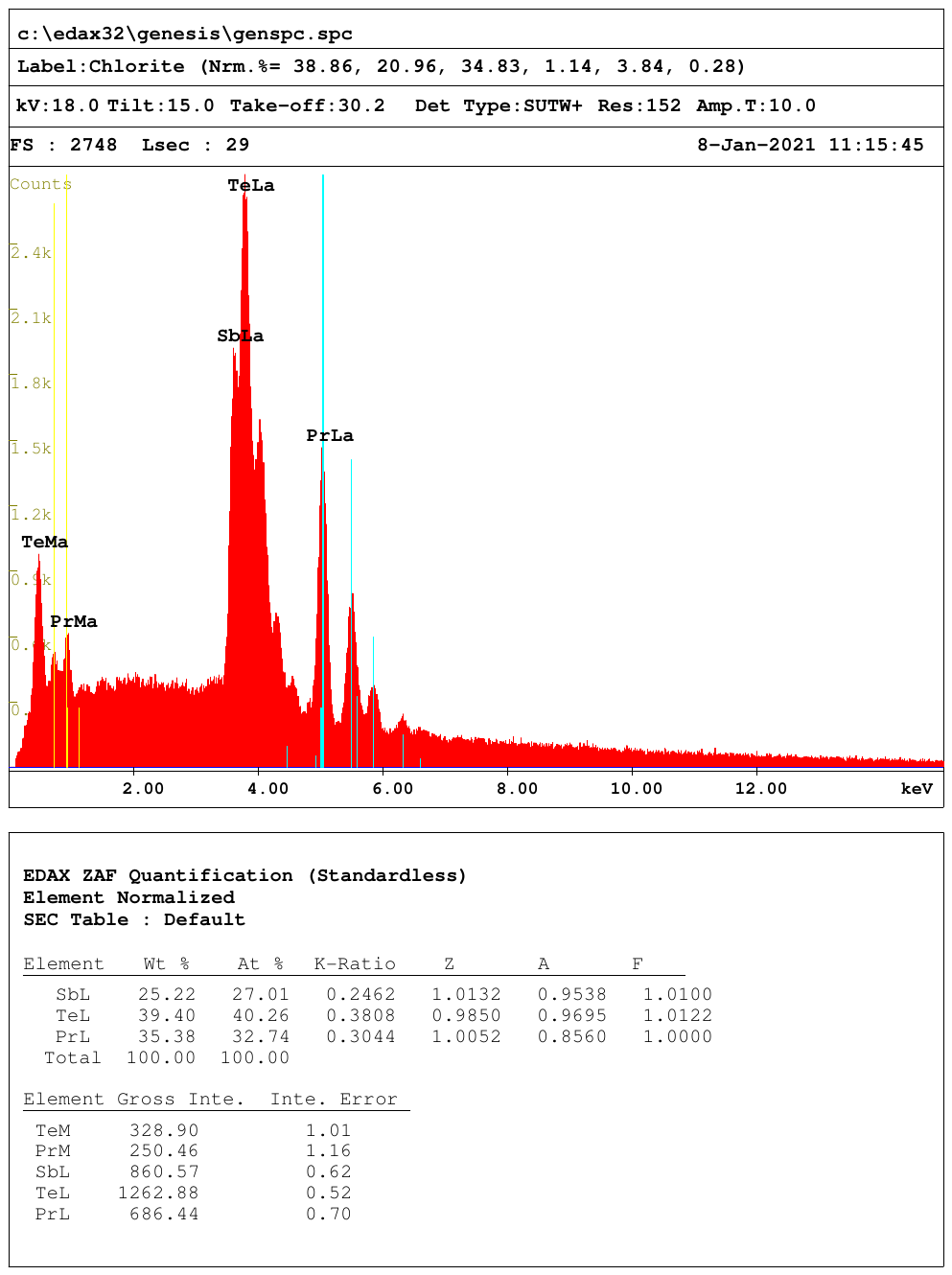}
\caption{Microprobe analysis of the chemical composition of PrSbTe single crystal used in the bulk property measurements.}
\label{SF1}
\end{figure*}

Synthesis of single-crystalline $\mathrm{PrSbTe}$ was carried out by employing chemical vapor transport method using Iodine as transport agent. Chemical composition and phase homogeneity of the prepared crystals were determined by energy-dispersive X-ray (EDX) analysis performed using a FEI scanning electron microscope equipped with an EDAX Genesis XM4 spectrometer. The EDX experiment indicated for the single crystal examined a chemical composition slightly deviating from the ideal 111 stoichiometry, namely Pr 32.74(6) at\%, Sb 27.01(9) at\%, Te 40.26(8) at\% (see Fig.~\ref{SF1}). Crystal structure examination of the single crystals was done using single crystal X-ray diffraction (XRD) carried out on an Oxford Diffraction X'calibur four-circle single-crystal X-ray diffractometer equipped with a CCD Atlas detector. Analysis of the XRD data collected on the crystal studied by EDX yielded the tetragonal lattice parameters $a$ = $b$ = 4.3594(3)$~\text{\AA}$ and $c$ = 9.3754(9)$~\text{\AA}$.

\subsubsection*{Magnetic and heat transport Measurements}
\begin{figure*} [h!]
\includegraphics[width=0.6\textwidth]{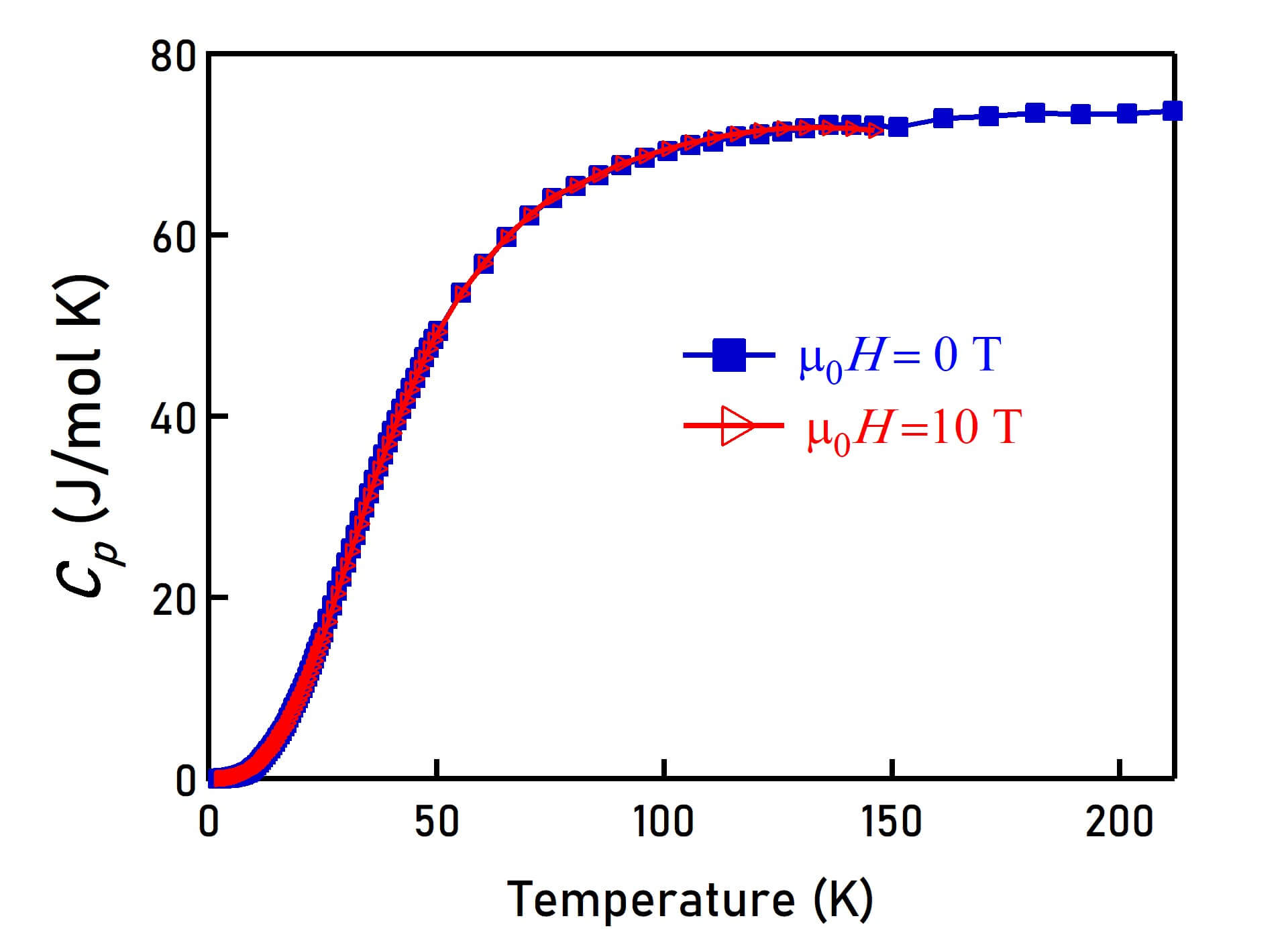}
\caption{Heat capacity measured in zero field and in a 10~T field.}
\label{SF2}
\end{figure*}

\begin{figure*} [h!]
\includegraphics[width=0.9\textwidth]{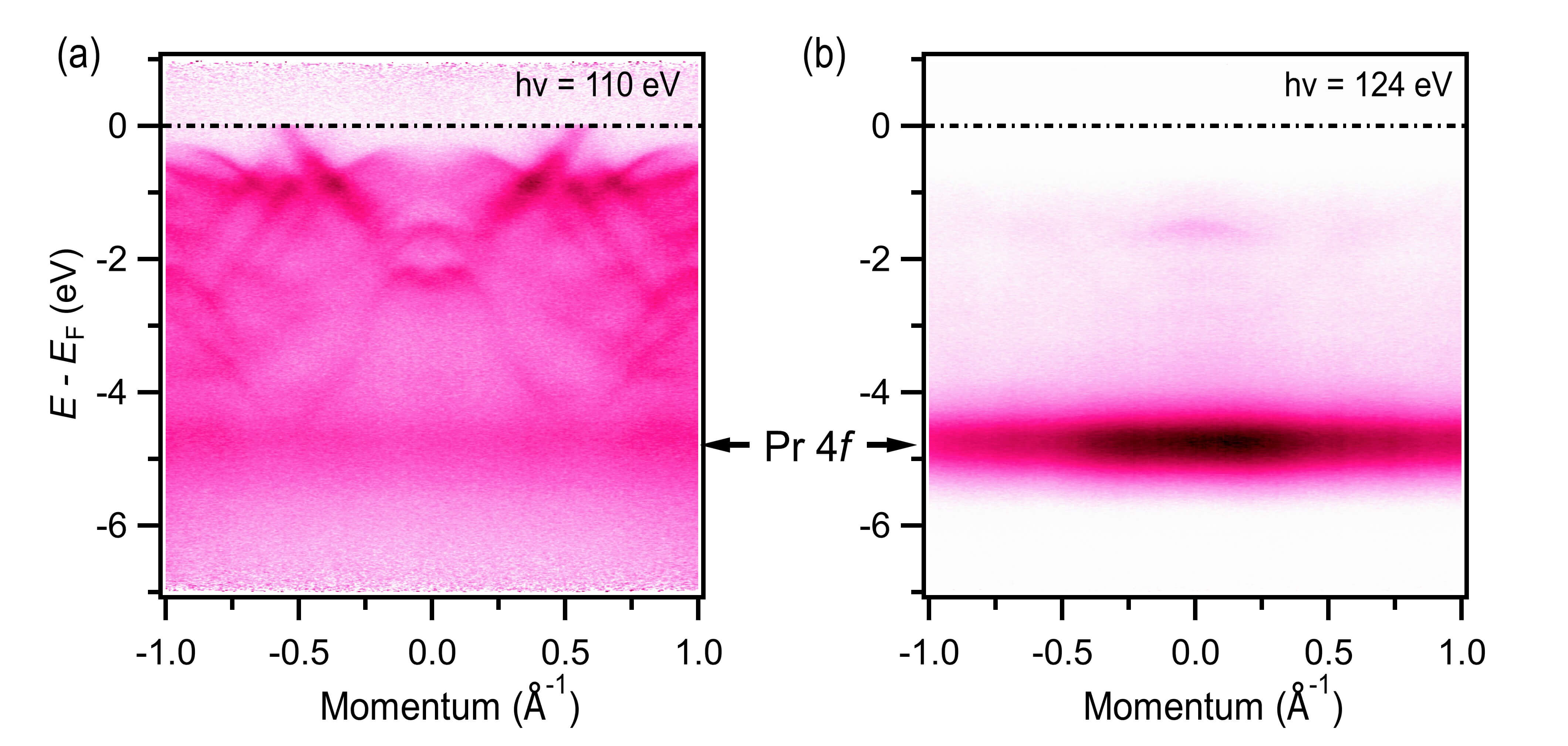}
\caption{Photoemission valence band spectrum showing Pr 4\textit{f} bands lying deep below the Fermi level.}
\label{SF3}
\end{figure*}

Magnetic property measurements were carried out using a Quantum Design MPMS-XL superconducting quantum interference device (SQUID) in a temperature interval 1.72-400~K and in magnetic fields up to 7~T.  Heat capacity was measured in in a Quantum Design PPMS-14 platform within the temperature interval 2-250~K and in magnetic fields up to 10~T. Both magnetic and heat capacity measurements show no signature of phase transitions down to $\sim$2 K,indicative of well localized 4\textit{f} states. This is in line with the result of photoemission valence band spectrum (Figure~\ref{SF3}), which shows the presence of 4\textit{f} bands deep below  the Fermi level. Application of large magnetic field up to 10~T does not bring any significant change in heat capacity (see Fig.~\ref{SF2}).

\subsubsection*{ARPES measurements}
Spectroscopic measurements of the electronic structure were performed employing the ARPES set ups at the Advanced Light Source (ALS) beamline 10.0.1.1 at a temperature of 15~K and at the Stanford Synchrotron Radiation Lightsource (SSRL) endstation 5-2 at a temperature of 7 K and with linear horizontal polarized light. The energy resolution of the experimental set up was set better than $\mathrm{0.02~eV}$. The samples were first mounted on copper posts using silver epoxy and then were loaded inside the ARPES chamber. The mounted samples were cleaved under ultra-high vacuum better than $\mathrm{10^{-10}~Torr}$ maintained in the ARPES chamber  before carrying out the measurements. 

\section*{S2. Computational Details}
The density-functional theory (DFT) \cite{Hohenberg1964S, Kohn1965S} based first-principles calculations were performed using projector augmented-wave (PAW) potentials \cite{Blochl1994S} implemented in the Vienna Ab initio Simulation Package ({\sc VASP}) \cite{Kresse1994S, Kresse1996S, Kresse1999S}. Calculations were made within the generalized gradient approximation (GGA) in the Perdew, Burke, and Ernzerhof (PBE) parameterization \cite{Perdew1996S}. The energy cutoff for the plane-wave expansion was set to $\mathrm{350~eV}$. The $4f$ states of $\mathrm{Pr}$ were treated within the DFT+$U$ scheme introduced by Liechtenstein  {\it et al.} \cite{Liechtenstein1995S} (with $\mathrm{U = 4.0~eV}$ and $\mathrm{J = 0.5~eV)}$.  \\

The electronic band structure was also evaluated within {\sc Quantum ESPRESSO}  \cite{Giannozzi2009S,Giannozzi2017S,Giannozzi2020S} with PAW GGA PBE pseudopotential included in {\sc PsLibrary} \cite{Corso2014S}. Exact results of the DFT calculations (with $\mathrm{Pr}$ $4f$ states as core states) were used to find the tight binding model in the basis of the maximally localized Wannier orbitals \cite{Marzhari2012S, Marzhari1997S, Souza2001S}. It was performed using the {\sc Wannier90} software \cite{Mostofi2008S, Mostofi2014S, Pizzi2020S}. The wannierization procedure (i.e., procedure of the tight binding model finding) is based on the exact DFT band structure. Within this procedure, storage of each wavefunction for all ${\bm k}$-points are necessary. Next, the Bloch wavefunction from DFT are used in the construction of the “rotation” and overlap matrices. More technical details can be found in the documentation of {\sc wannier90} or Ref. \cite{Marzhari2012S}. Finally, the found $22$-orbital tight binding model of the PrSbTe was used for the investigation of the surface Green's function for semi-infinite system \cite{Sancho1985S} using {\sc WannierTools} \cite{Wu2018S} software. Within this calculation, the tight binding models of the bulk system is used to perform the properties of the system with “edge” (e.g. surface). More detailed theoretical description can be found in Ref. \cite{Sancho1985S}. 

\section*{S3. Wide valence band structure}
In Fig. \ref{SF3}, we present the wide valence band structure measured with photon energies of 110 eV (off $4d-4f$ resonance) and 124 eV (near $4d-4f$ resonance). From both measurements, we can see that the Pr $4f$ bands lie deep below the Fermi level. Near the resonance, the intensity of Pr $4f$ bands sharply increases dominating the spectral intensity of other bands.

\section*{S3. Slab Calculation}
\begin{figure*} [h]
\includegraphics[width=0.9\textwidth]{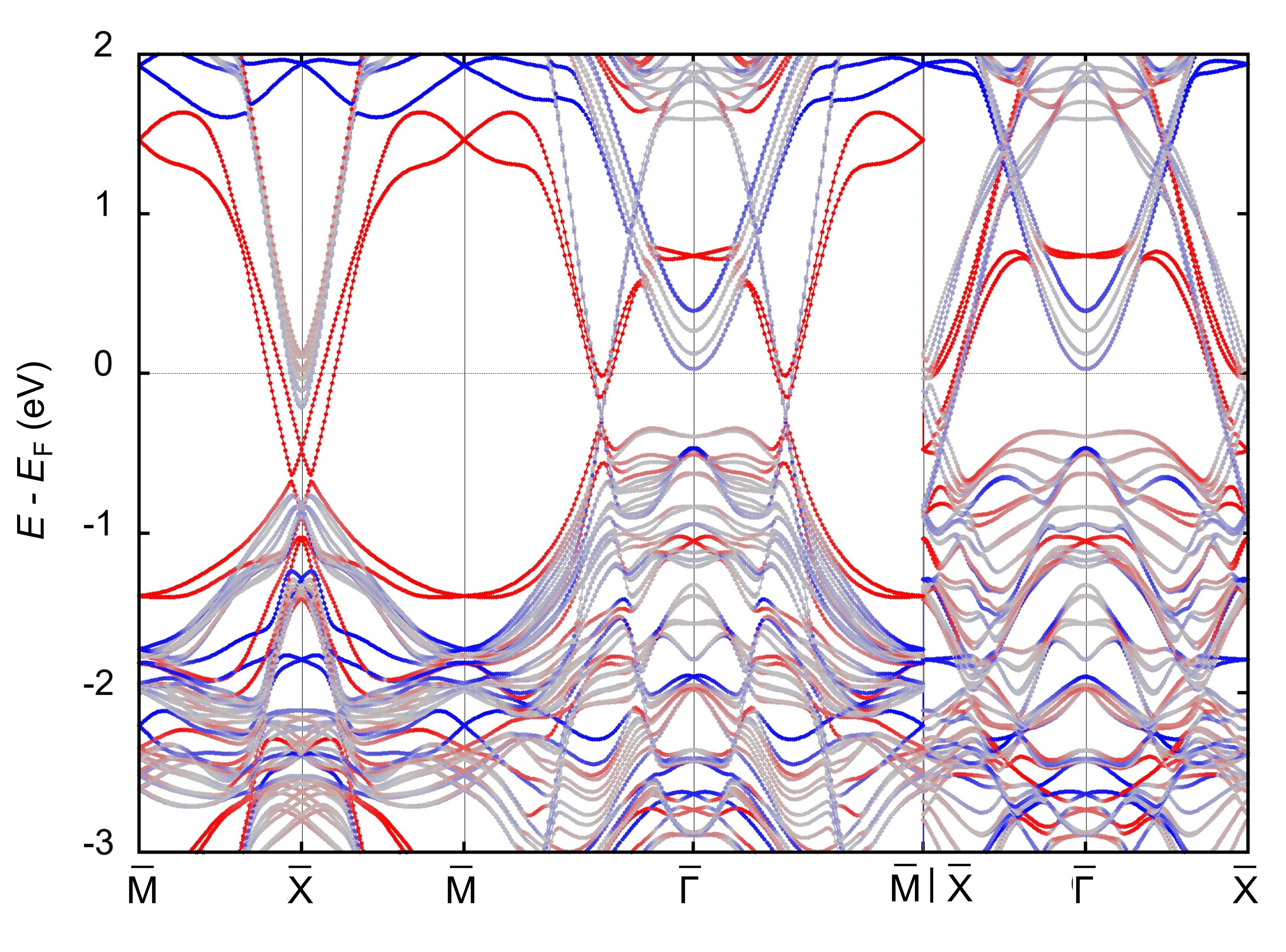}
\caption{Slab calculations along the $\mathrm{\overline{M}} - \mathrm{\overline{X}} - \mathrm{\overline{M}}$, $\mathrm{\overline{M}} - \mathrm{\overline{\Gamma}} - \mathrm{\overline{M}}$, and $\mathrm{\overline{X}} - \mathrm{\overline{\Gamma}} - \mathrm{\overline{X}}$ directions. The grey, red, and blue colors represent bulk, $\mathrm{Sb}$ surface, and  $\mathrm{PrTe}$ plane terminated surface bands, respectively.}
\label{SF4}
\end{figure*}

In Fig. \ref{SF4}, we present the slab calculations along the $\mathrm{\overline{M}} - \mathrm{\overline{X}} - \mathrm{\overline{M}}$, $\mathrm{\overline{M}} - \mathrm{\overline{\Gamma}} - \mathrm{\overline{M}}$, and $\mathrm{\overline{X}} - \mathrm{\overline{\Gamma}} - \mathrm{\overline{X}}$ directions. The grey bands represent the surface projected bulk bands, red bands are surface bands for $\mathrm{Sb}$ square net termination, and the blue bands are the surface states with $\mathrm{PrTe}$ plane termination. The comparison with our experimental data shows that the termination of the sample is on the $\mathrm{PrTe}$ plane.

\begin{figure*} [h]
\includegraphics[width=1\textwidth]{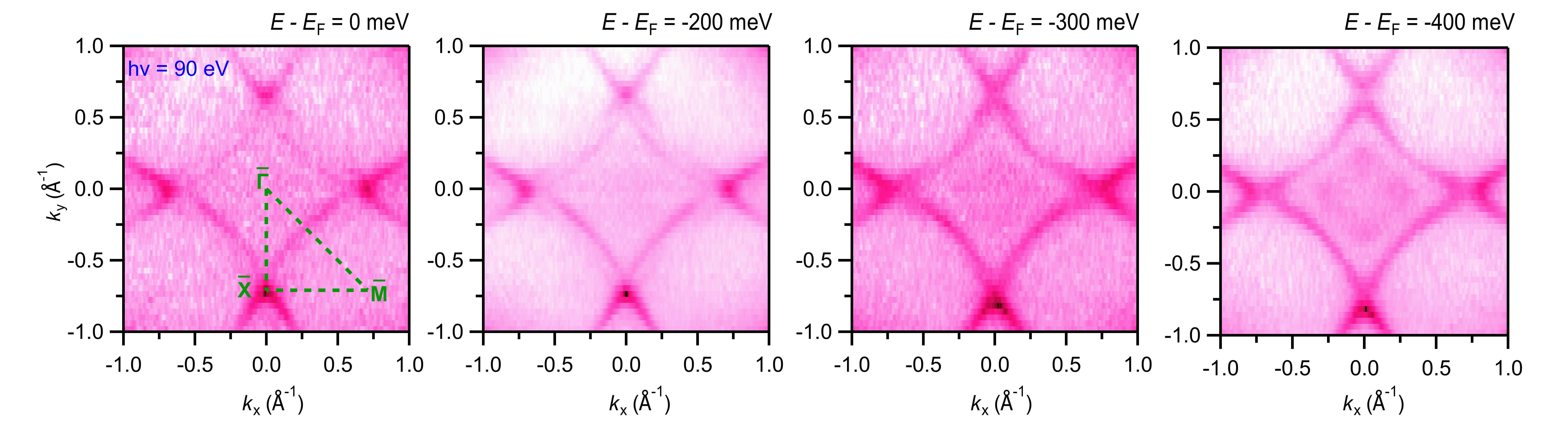}
\caption{Fermi surface and energy contours taken with a photon energy of 90 eV at SSRL 5-2 beamline at a temperature of 7 K with linear horizontal polarized light.}
\label{SF5}
\end{figure*}

\begin{figure*} [h]
\includegraphics[width=1\textwidth]{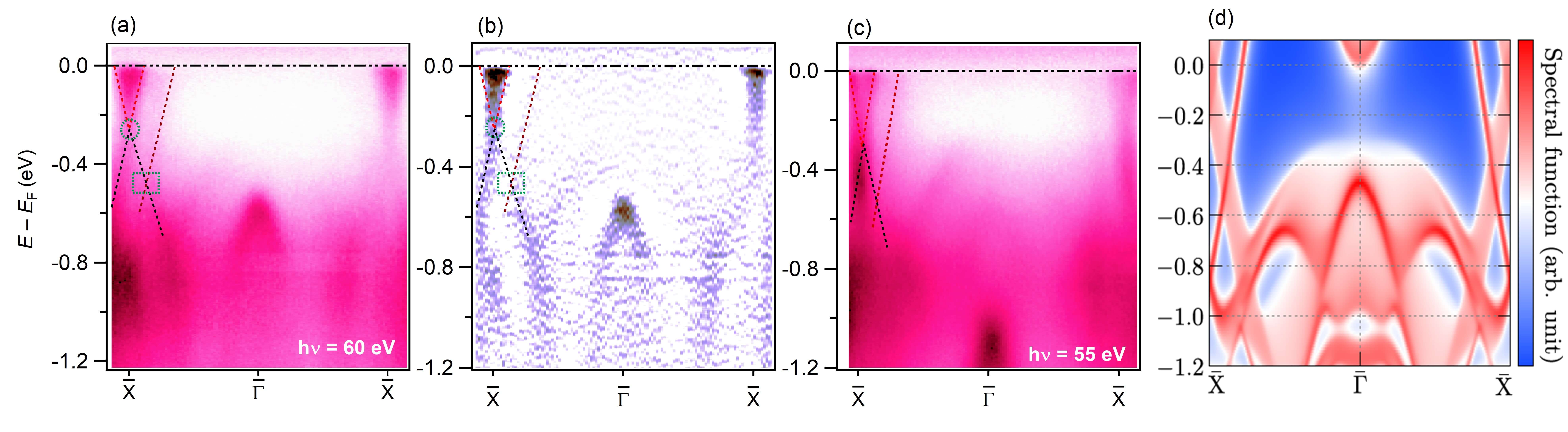}
\caption{(a) Dispersion map along $\mathrm{\overline{X}} - \mathrm{\overline{\Gamma}}$ measured using photon energies of 60 eV. (b)Second derivative of (a). (c) Dispersion map along $\mathrm{\overline{X}} - \mathrm{\overline{\Gamma}}$ measured using photon energies of 55~eV. (d) Surface spectrum along $\mathrm{\overline{X}} - \mathrm{\overline{\Gamma}}$ obtained from theoretical calculations. Experimental data were collected at the ALS 10.0.1.1 beamline at a temperature of 15 K.}
\label{SF6}
\end{figure*}

\begin{figure*} [h]
\includegraphics[width=0.85\textwidth]{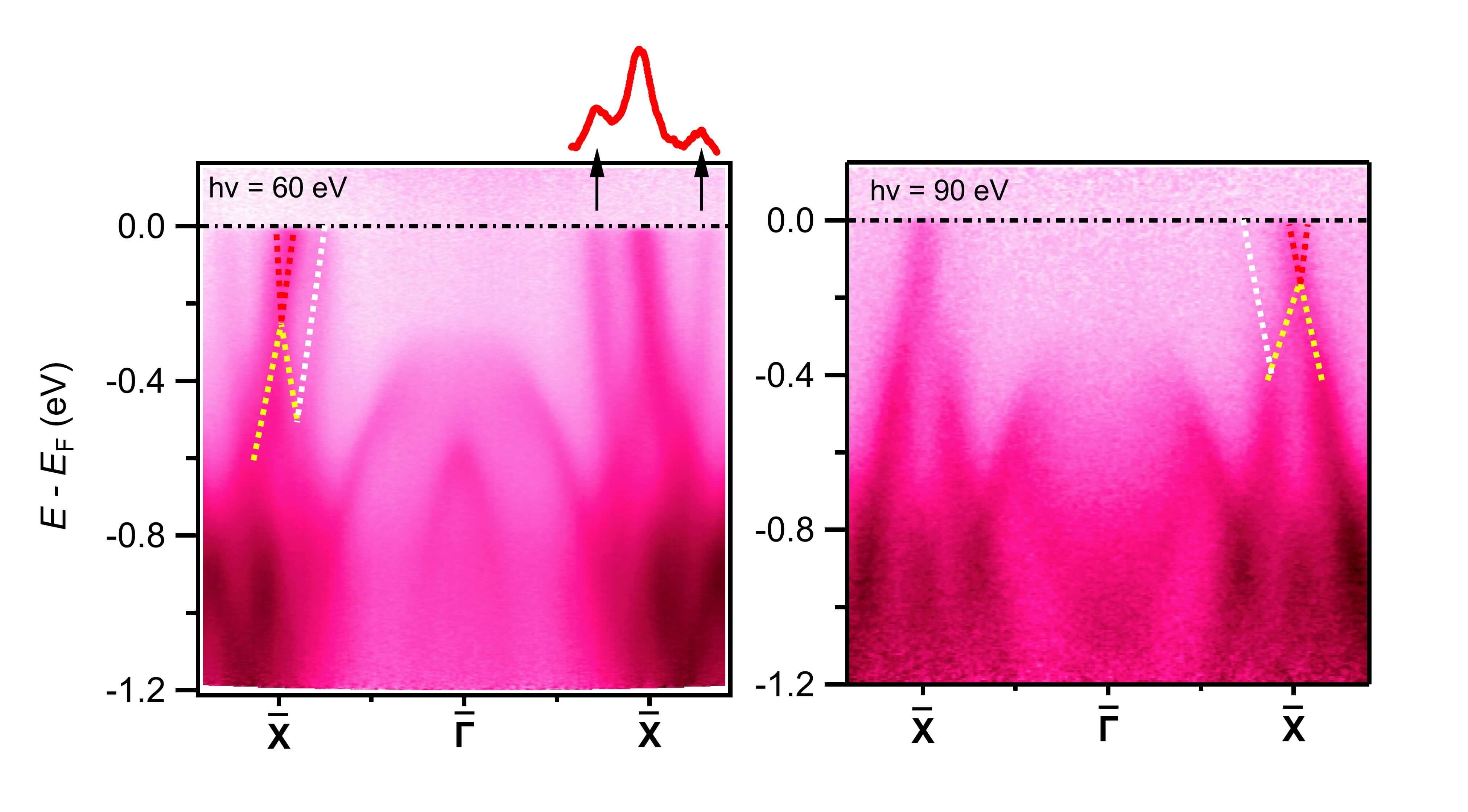}
\caption{Dispersion map along $\mathrm{\overline{X}} - \mathrm{\overline{\Gamma}}$ at photon energies of 60 eV (left) and 90 eV (right). Left plot shows momentum distribution curve integrated within (0, -0.05~eV) binding energy window. Data were collected at the SSRL 5-2 beamline at a temperature of 7 K with linear horizontal polarized light.}
\label{SF7}
\end{figure*}


\begin{thebibliography}{50}



\bibitem{Hasan2010} M. Z. Hasan and C. L. Kane, Colloquium: Topological insulators, \href{https://doi.org/10.1103/RevModPhys.82.3045}{Rev. Mod. Phys. \textbf{82}, 3045 (2010)}.

\bibitem{Qi2011} X.-L. Qi and S.-C. Zhang, Topological insulators and superconductors, \href{https://doi.org/10.1103/RevModPhys.83.1057}{Rev. Mod. Phys. \textbf{83}, 1057 (2011)}.

\bibitem{Armitage2018} N. P. Armitage, E. J. Mele, and A. Vishwanath, Weyl and Dirac semimetals in three-dimensional solids, \href{https://doi.org/10.1103/RevModPhys.90.015001}{Rev. Mod. Phys. \textbf{90}, 015001 (2018)}.

\bibitem{Sato2017} M. Sato and Y. Ando, Topological superconductors: a review, \href{https://doi.org/10.1088/1361-6633/aa6ac7}{Rep. Prog. Phys. \textbf{80}, 076501 (2017)}.

\bibitem{Liu2014} Z. K. Liu, B. Zhou, Y. Zhang, Z. J. Wang, H. M. Weng, D. Prabhakaran, S. K. Mo, Z. X. Shen, Z. Fang, X. Dai, Z. Hussain, and Y. L. Chen, Discovery of a Three-Dimensional Topological Dirac Semimetal, $\mathrm{Na}_3\mathrm{Bi}$, \href{https://doi.org/10.1126/science.124508}{Science \textbf{343}, 864 (2014)}.

\bibitem{Neupane2014} M. Neupane, S.-Y. Xu, R. Sankar, N. Alidoust, G. Bian, C. Liu, I. Belopolski, T.-R. Chang, H.-T. Jeng, H. Lin, A. Bansil, F. Chou, and M. Z. Hasan, Observation of a three-dimensional topological Dirac semimetal phase in high-mobility $\mathrm{Cd}_3\mathrm{As}_2$, \href{https://doi.org/10.1038/ncomms4786}{Nat. Commun. \textbf{5}, 3786 (2014)}.


\bibitem{Xu2015} S.-Y. Xu, I. Belopolski, N. Alidoust, M. Neupane, G. Bian, C. Zhang, R. Sankar, G. Chang, Z. Yuan, C.-C. Lee, S.-M. Huang, H. Zheng, J. Ma, D. S. Sanchez, B. Wang, A. Bansil, F. Chou, P. P. Shibayev, H. Lin, S. Jia, and M. Z. Hasan, Discovery of a Weyl fermion semimetal and topological Fermi arcs, \href{https://doi.org/10.1126/science.aaa929}{Science \textbf{349}, 613 (2015)}.

\bibitem{Lv2015}	B. Q. Lv, H. M. Weng, B. B. Fu, X. P. Wang, H. Miao, J. Ma, P. Richard, X. C. Huang, L. X. Zhao, G. F. Chen, Z. Fang, X. Dai, T. Qian, and H. Ding, Experimental Discovery of Weyl Semimetal $\mathrm{TaAs}$, \href{https://doi.org/10.1103/PhysRevX.5.031013}{Phys. Rev. X \textbf{5}, 031013 (2015)}.

\bibitem{Soluyanov2015} A. A. Soluyanov, D. Gresch, Z. Wang, Q. Wu, M. Troyer, X. Dai, and B. A. Bernevig, Type-II Weyl semimetals, \href{https://doi.org/10.1038/nature15768}{Nature \textbf{527}, 495 (2015)}.

\bibitem{Yan2017} B. Yan and C. Felser, Topological Materials: Weyl Semimetals, \href{https://doi.org/10.1146/annurev-conmatphys-031016-025458}{Annu. Rev. Condens. Matter Phys. \textbf{8}, 337 (2017)}.



\bibitem{Burkov2011} A. A. Burkov, M. D. Hook, and L. Balents, Topological nodal semimetals, \href{https://doi.org/10.1103/PhysRevB.84.235126}{Phys. Rev. B \textbf{84}, 235126 (2011)}.

\bibitem{Bian2016} G. Bian, T.-R. Chang, R. Sankar, S.-Y. Xu, H. Zheng, T. Neupert, C.-K. Chiu, S.-M. Huang, G. Chang, I. Belopolski, D. S. Sanchez, M. Neupane, N. Alidoust, C. Liu, B. Wang, C.-C. Lee, H.-T. Jeng, C. Zhang, Z. Yuan, S. Jia, A. Bansil, F. Chou, H. Lin, and M. Z. Hasan, Topological nodal-line fermions in spin-orbit metal $\mathrm{Pb}\mathrm{Ta}\mathrm{Se}_2$, \href{https://doi.org/10.1038/ncomms10556}{Nat. Commun. \textbf{7}, 10556 (2016)}.

\bibitem{Neupane2016} M. Neupane, I. Belopolski, M. M. Hosen, D. S. Sanchez, R. Sankar, M. Szlawska, S.-Y. Xu, K. Dimitri, N. Dhakal, P. Maldonado, P. M. Oppeneer, D. Kaczorowski, F. Chou, M. Z. Hasan, and T. Durakiewicz, Observation of topological nodal fermion semimetal phase in $\mathrm{ZrSiS}$, \href{https://doi.org/10.1103/PhysRevB.93.201104}{Phys. Rev. B \textbf{93}, 201104 (2016)}.

\bibitem{Schoop2016} L. M. Schoop, M. N. Ali, C. Straßer, A. Topp, A. Varykhalov, D. Marchenko, V. Duppel, S. S. P. Parkin, B. V. Lotsch, and C. R. Ast, Dirac cone protected by non-symmorphic symmetry and three-dimensional Dirac line node in $\mathrm{ZrSiS}$, \href{https://doi.org/10.1038/ncomms11696}{Nat. Commun. \textbf{7}, 11696 (2016)}.

\bibitem{Yang2018} S.-Y. Yang, H. Yang, E. Derunova, S. S. P.Parkin, B. Yan, and M. N. Ali, Symmetry demanded topological nodal-line materials, \href{https://doi.org/10.1080/23746149.2017.1414631}{Adv. Phys. X \textbf{3}, 1414631 (2018)}.

\bibitem{Topp2016} A. Topp, J. M. Lippmann, A. Varykhalov, V. Duppel, B. V. Lotsch, C. R. Ast, and L. M. Schoop, Non-symmorphic band degeneracy at the Fermi level in $\mathrm{ZrSiTe}$, \href{https://doi.org/10.1088/1367-2630/aa4f65}{New J. Phys. \textbf{18}, 125014 (2016)}.

\bibitem{Hu2016} J. Hu, Z. Tang, J. Liu, X. Liu, Y. Zhu, D. Graf, K. Myhro, S. Tran, C. N. Lau, J. Wei, and Z. Mao, Evidence of Topological Nodal-Line Fermions in $\mathrm{ZrSiSe}$ and $\mathrm{ZrSiTe}$, \href{https://doi.org/10.1103/PhysRevLett.117.016602}{Phys. Rev. Lett. \textbf{117}, 016602 (2016)}.

\bibitem{Takane2016} D. Takane, Z. Wang, S. Souma, K. Nakayama, C. X. Trang, T. Sato, T. Takahashi, and Y. Ando, Dirac-node arc in the topological line-node semimetal $\mathrm{HfSiS}$, \href{https://doi.org/10.1103/PhysRevB.94.121108}{Phys. Rev. B \textbf{94}, 121108 (2016)}.

\bibitem{Lou2016} R. Lou, J. Z. Ma, Q. N. Xu, B. B. Fu, L. Y. Kong, Y. G. Shi, P. Richard, H. M. Weng, Z. Fang, S. S. Sun, Q. Wang, H. C. Lei, T. Qian, H. Ding, and S. C. Wang, Emergence of topological bands on the surface of $\mathrm{ZrSnTe}$ crystal, \href{https://doi.org/10.1103/PhysRevB.93.241104}{Phys. Rev. B \textbf{93}, 241104 (2016)}.

\bibitem{Hosen2017} M. M. Hosen, K. Dimitri, I. Belopolski, P. Maldonado, R. Sankar, N. Dhakal, G. Dhakal, T. Cole, P. M. Oppeneer, D. Kaczorowski, F. Chou, M. Z. Hasan, T. Durakiewicz, and M. Neupane, Tunability of the topological nodal-line semimetal phase in $\mathrm{ZrSi}X$-type materials ($X=\mathrm{S}, \mathrm{Se}, \mathrm{Te}$), \href{https://doi.org/10.1103/PhysRevB.95.161101}{Phys. Rev. B \textbf{95}, 161101 (2017)}.

\bibitem{Chen2017} C. Chen, X. Xu, J. Jiang, S. C. Wu, Y. P. Qi, L. X. Yang, M. X. Wang, Y. Sun, N. B. M. Schröter, H. F. Yang, L. M. Schoop, Y. Y. Lv, J. Zhou, Y. B. Chen, S. H. Yao, M. H. Lu, Y. F. Chen, C. Felser, B. H. Yan, Z. K. Liu, and Y. L. Chen, Dirac line nodes and effect of spin-orbit coupling in the nonsymmorphic critical semimetals $M\mathrm{SiS}\phantom{\rule{0.16em}{0ex}}(M=\mathrm{Hf},\phantom{\rule{0.16em}{0ex}}\mathrm{Zr})$, \href{https://doi.org/10.1103/PhysRevB.95.125126}{Phys. Rev. B \textbf{95}, 125126 (2017)}.

\bibitem{Hosen2018} M. M. Hosen, K. Dimitri, A. Aperis, P. Maldonado, I. Belopolski, G. Dhakal, F. Kabir, C. Sims, M. Z. Hasan, D. Kaczorowski, T. Durakiewicz, P. M. Oppeneer, and M. Neupane, Observation of gapless Dirac surface states in $\mathrm{ZrGeTe}$, \href{https://doi.org/10.1103/PhysRevB.97.121103}{Phys. Rev. B \textbf{97}, 121103 (2018)}.

\bibitem{Fu2019} B. B. Fu, C. J. Yi, T. T. Zhang, M. Caputo, J. Z. Ma, X. Gao, B. Q. Lv, L. Y. Kong, Y. B. Huang, P. Richard, M. Shi, V. N. Strocov, C. Fang, H. M. Weng, Y. G. Shi, T. Qian, and H. Ding, Dirac nodal surfaces and nodal lines in $\mathrm{ZrSiS}$, \href{https://doi.org/10.1126/sciadv.aau6459}{Sci. Adv. \textbf{5}, eaau6459 (2019)}.

\bibitem{Chen2017Ce} K. W. Chen, Y. Lai, Y. C. Chiu, S. Steven, T. Besara, D. Graf, T. Siegrist, T. E. Albrecht-Schmitt, L. Balicas, and R. E. Baumbach, Possible devil's staircase in the Kondo lattice $\mathrm{CeSbTe}$, \href{https://doi.org/10.1103/PhysRevB.96.014421}{Phys. Rev. B \textbf{96}, 014421 (2017)}.

\bibitem{Hosen2018Gd} M. M. Hosen, G. Dhakal, K. Dimitri, P. Maldonado, A. Aperis, F. Kabir, C. Sims, P. Riseborough, P. M. Oppeneer, D. Kaczorowski, T. Durakiewicz, and M. Neupane, Discovery of topological nodal-line fermionic phase in a magnetic material $\mathrm{GdSbTe}$, \href{https://doi.org/10.1038/s41598-018-31296-7}{Sci. Rep. \textbf{8}, 13283 (2018)}.

\bibitem{Schoop2018}  L. M. Schoop, A. Topp, J. Lippmann, F. Orlandi, L. Müchler, M. G. Vergniory, Y. Sun, A. W. Rost, V. Duppel, M. Krivenkov, S. Sheoran, P. Manuel, A. Varykhalov, B. Yan, R. K. Kremer, C. R. Ast, and B. V. Lotsch, Tunable Weyl and Dirac states in the nonsymmorphic compound $\mathrm{CeSbTe}$, \href{https://doi.org/10.1126/sciadv.aar2317}{Sci. Adv. \textbf{4}, eaar2317 (2018)}.

\bibitem{Sankar2019} R. Sankar, I. P. Muthuselvam, K. R. Babu, G. S. Murugan, K. Rajagopal, R. Kumar, T.-C. Wu, C.-Y. Wen, W.-L. Lee, G.-Y. Guo, and F.-C. Chou, Crystal Growth and Magnetic Properties of Topological Nodal-Line Semimetal $\mathrm{GdSbTe}$ with Antiferromagnetic Spin Ordering, \href{https://doi.org/10.1021/acs.inorgchem.9b01698}{Inorg. Chem. \textbf{58}, 11730 (2019)}.

\bibitem{Yang2020} M. Yang, Y. Qian, D. Yan, Y. Li, Y. Song, Z. Wang, C. Yi, H. L. Feng, H. Weng, and Y. Shi, Magnetic and electronic properties of a topological nodal line semimetal candidate: $\mathrm{HoSbTe}$, \href{https://doi.org/10.1103/PhysRevMaterials.4.094203}{Phys. Rev. Mater. \textbf{4}, 094203 (2020)}.

\bibitem{Lv2019}  B. Lv, J. Chen, L. Qiao, J. Ma, X. Yang, M. Li, M. Wang, Q. Tao, and Z.-A. Xu, Magnetic and transport properties of low-carrier-density Kondo semimetal $\mathrm{CeSbTe}$, \href{https://orcid.org/0000-0001-9290-2762}{J. Phys. Condens. Mater. \textbf{31}, 355601 (2019)}.

\bibitem{Pandey2020} K. Pandey, R. Basnet, A. Wegner, G. Acharya, M. R. U. Nabi, J. Liu, J. Wang, Y. K. Takahashi, B. Da, and J. Hu, Electronic and magnetic properties of the topological semimetal candidate $\mathrm{NdSbTe}$, \href{https://doi.org/10.1103/PhysRevB.101.235161}{Phys. Rev. B 101, 235161 (2020)}.

\bibitem{Pandey2021}  K. Pandey, D. Mondal, J. W. Villanova, J. Roll, R. Basnet, A. Wegner, G. Acharya, M. R. U. Nabi, B. Ghosh, J. Fujii, J. Wang, B. Da, A. Agarwal, I. Vobornik, A. Politano, S. Barraza-Lopez, and J. Hu, Magnetic Topological Semimetal Phase with Electronic Correlation Enhancement in SmSbTe, \href{https://doi.org/10.1002/qute.202100063}{Adv. Quantum Technol. \textbf{4}, 2100063 (2021)}.

\bibitem{Plokhikh2022} I. Plokhikh, V. Pomjakushin, D. J. Gawryluk, O. Zaharko, and E. Pomjakushina, Competing Magnetic Phases in $Ln\mathrm{Sb}\mathrm{Te}$ ($Ln$ = $\mathrm{Ho}$ and $\mathrm{Tb}$), \href{https://doi.org/10.1021/acs.inorgchem.2c01711}{Inorg. Chem. \textbf{61}, 11399 (2022)}.

\bibitem{Gao2022} F. Gao, J. Huang, W. Ren, M. Li, H. Wang, T. Yang, B. Li, and Z. Zhang, Magnetic and transport properties of the topological compound $\mathrm{DySbTe}$, \href{https://doi.org/10.1103/PhysRevB.105.214434}{Phys. Rev. B \textbf{105}, 214434 (2022)}.

\bibitem{Regmi2023} S. Regmi, R. Smith, A. P. Sakhya, M. Sprague, M. I. Mondal, I. Bin Elius, N. Valadez, A. Ptok, D. Kaczorowski, and M. Nepane, Observation of gapless nodal-line states in $\mathrm{NdSbTe}$, \href{https://doi.org/10.1103/PhysRevMaterials.7.044202}{Phys. Rev. Materials \textbf{7}, 044202 (2023)}.

\bibitem{Gao2023} F. Gao, J. Huang, W. Ren, H. Wu, M. An, X. Wu, L. Zhang, T. Yang, A. Wang, Y. Chai, X. Zhao, T. Yang, B. Li, and Z. Zhang, Magnetic and Magnetotransport Properties of the Magnetic Topological Nodal-Line Semimetal $\mathrm{TbSbTe}$, \href{https://doi.org/10.1002/qute.202200163}{Adv. Quantum Technol. 2200163 (2023)}.

\bibitem{Li2021} P. Li, B. Lv, Y. Fang, W. Guo, Z. Wu, Y. Wu, D. Shen, Y. Nie, L. Petaccia, C. Cao, Z.-A. Xu, and Y. Liu, Charge density wave and weak Kondo effect in a Dirac semimetal $\mathrm{CeSbTe}$, \href{https://doi.org/10.1007/s11433-020-1642-2}{Sci. China Phys. Mech. Astron. \textbf{64}, 237412 (2021)}.

\bibitem{Wang2021} Y. Wang, Y. Qian, M. Yang, H. Chen, C. Li, Z. Tan, Y. Cai, W. Zhao, S. Gao, Y. Feng, S. Kumar, E. F. Schwier, L. Zhao, H. Weng, Y. Shi, G. Wang, Y. Song, Y. Huang, K. Shimada, Z. Xu, X. J. Zhou, and G. Liu, Spectroscopic evidence for the realization of a genuine topological nodal-line semimetal in $\mathrm{LaSbTe}$, \href{https://doi.org/10.1103/PhysRevB.103.125131}{Phys. Rev. B \textbf{103}, 125131 (2021)}.

\bibitem{Regmi2022} S. Regmi, G. Dhakal, F. C. Kabeer, N. Harrison, F. Kabir, A. P. Sakhya, K. Gofryk, D. Kaczorowski, P. M. Oppeneer, and M. Neupane, Observation of multiple nodal lines in $\mathrm{SmSbTe}$, \href{https://doi.org/10.1103/PhysRevMaterials.6.L031201}{Phys. Rev. Materials \textbf{6}, L031201 (2022)}.

\bibitem{Yue2020} S. Yue, Y. Qian, M. Yang, D. Geng, C. Yi, S. Kumar, K. Shimada, P. Cheng, L. Chen, Z. Wang, H. Weng, Y. Shi, K. Wu, and B. Feng, Topological electronic structure in the antiferromagnet $\mathrm{HoSbTe}$, \href{https://doi.org/10.1103/PhysRevB.102.155109}{Phys. Rev. B \textbf{102}, 155109 (2020)}.

\bibitem{Shumiya2022} N. Shumiya, J.-X. Yin, G. Chang, M. Yang, S. Mardanya, T.-R. Chang, H. Lin, M. S. Hossain, Y.-X. Jiang, T. A. Cochran, Q. Zhang, X. P. Yang, Y. Shi, and M. Z. Hasan, Evidence for electronic signature of a magnetic transition in the topological magnet $\mathrm{HoSbTe}$, \href{https://doi.org/10.1103/PhysRevB.106.035151}{Phys. Rev. B \textbf{106}, 035151 (2022)}.

\bibitem{Topp2019} A. Topp, M. G. Vergniory, M. Krivenkov, A. Varykhalov, F. Rodolakis, J. L. McChesney, B. V. Lotsch, C. R. Ast, and L. M. Schoop, The effect of spin-orbit coupling on nonsymmorphic square-net compounds, \href{https://doi.org/10.1016/j.jpcs.2017.12.035}{J.  Phys. Chem.  Solids \textbf{128}, 296 (2019)}.

\bibitem{Hohenberg1964} P. Hohenber and W. Kohn,  Inhomogeneous Electron Gas,  \href{https://doi.org/10.1103/PhysRev.136.B864}{Phys. Rev. \textbf{136,} B864 (1964)}.

\bibitem{Kohn1965} W. Kohn and L. J. Sham, Self-Consistent Equations Including Exchange and Correlation Effects, \href{https://doi.org/10.1103/PhysRev.140.A1133}{Phys. Rev. \textbf{140}, A1133 (1965)}.

\bibitem{Kresse1994} G. Kresse and J. Hafner, Ab initio molecular-dynamics simulation of the liquid-metal--amorphous-semiconductor transition in germanium, \href{https://doi.org/10.1103/PhysRevB.49.14251}{Phys. Rev. B \textbf{49}, 14251 (1994)}.

\bibitem{Kresse1996} G. Kresse and J. Furthmüller, Efficient iterative schemes for ab initio total-energy calculations using a plane-wave basis set, \href{https://doi.org/10.1103/PhysRevB.54.11169}{Phys. Rev. B \textbf{54,} 11169 (1996)}.

\bibitem{Kresse1999} G. Kresse and D. Joubert, From ultrasoft pseudopotentials to the projector augmented-wave method, \href{https://doi.org/10.1103/PhysRevB.59.1758}{Phys. Rev. B \textbf{59}, 1758 (1999)}.

\bibitem{Giannozzi2009} P. Giannozzi, S. Baroni, N. Bonini, M. Calandra, R. Car, C. Cavazzoni, D. Ceresoli, G. L. Chiarotti, M. Cococcioni, I. Dabo, A. Dal Corso, S. de Gironcoli, S. Fabris, G. Fratesi, R. Gebauer, U. Gerstmann, C. Gougoussis, A. Kokalj, M. Lazzeri, L. Martin-Samos, N. Marzari, F. Mauri, R. Mazzarello, S. Paolini, A. Pasquarello, L. Paulatto, C. Sbraccia, S. Scandolo, G. Sclauzero, A. P. Seitsonen, A. Smogunov, P. Umari, and R. M. Wentzcovitch, QUANTUM ESPRESSO: a modular and open-source software project for quantum simulations of materials, \href{https://doi.org/10.1088/0953-8984/21/39/395502}{J. Phys. Condens. Mater. \textbf{21}, 395502 (2009)}.

\bibitem{Giannozzi2017} P. Giannozzi, O. Andreussi, T. Brumme, O. Bunau, M. Buongiorno Nardelli, M. Calandra, R. Car, C. Cavazzoni, D. Ceresoli, M. Cococcioni, N. Colonna, I. Carnimeo, A. Dal Corso, S. de Gironcoli, P. Delugas, R. A. DiStasio, A. Ferretti, A. Floris, G. Fratesi, G. Fugallo, R. Gebauer, U. Gerstmann, F. Giustino, T. Gorni, J. Jia, M. Kawamura, H. Y. Ko, A. Kokalj, E. Küçükbenli, M. Lazzeri, M. Marsili, N. Marzari, F. Mauri, N. L. Nguyen, H. V. Nguyen, A. Otero-de-la-Roza, L. Paulatto, S. Poncé, D. Rocca, R. Sabatini, B. Santra, M. Schlipf, A. P. Seitsonen, A. Smogunov, I. Timrov, T. Thonhauser, P. Umari, N. Vast, X. Wu, and S. Baroni, Advanced capabilities for materials modelling with Quantum ESPRESSO, \href{https://doi.org/10.1088/1361-648X/aa8f79}{J. Phys. Condens. Mater. \textbf{29}, 465901 (2017)}.

\bibitem{Giannozzi2020} P. Giannozzi, O. Baseggio, P. Bonfà, D. Brunato, R. Car, I. Carnimeo, C. Cavazzoni, S. de Gironcoli, P. Delugas, F. Ferrari Ruffino, A. Ferretti, N. Marzari, I. Timrov, A. Urru, and S. Baroni, Quantum ESPRESSO toward the exascale, \href{https://doi.org/10.1063/5.0005082}{J. Chem. Phys.  \textbf{152}, 154105 (2020)}.

\bibitem{Blochl1994} P. E. Blöchl, Projector augmented-wave method,  \href{https://doi.org/10.1103/PhysRevB.50.17953}{Phys. Rev. B \textbf{50,} 17953 (1994)}.

\bibitem{SM} Please see Supplemental Material, containing references \cite{Hohenberg1964, Kohn1965, Blochl1994,Kresse1994, Kresse1996, Kresse1999, Perdew1996, Liechtenstein1995, Giannozzi2009, Giannozzi2017, Giannozzi2020, Corso2014, Marzhari2012, Marzhari1997, Souza2001, Mostofi2008, Mostofi2014, Pizzi2020, Sancho1985, Wu2018}, for sample characterization, slab calculation, and additional ARPES results.

\bibitem{Plokhikh2023} I. Plokhikh, V. Pomjakushin, D. J. Gawryluk, O. Zaharko, and E. Pomjakushina, On the magnetic structures of 1:1:1 stoichiometric topological phases $Ln\mathrm{SbTe}$ ($Ln$=$\mathrm{Pr}$, $\mathrm{Nd}$, $\mathrm{Dy}$, and $\mathrm{Er}$), \href{https://doi.org/10.1016/j.jmmm.2023.171009}{J. Magn. Magn. Mater. \textbf{583,} 171009 (2023)}.

\bibitem{Perdew1996} J. P. Perdew, K. Burke, and M. Ernzerhof, Generalized Gradient Approximation Made Simple, \href{https://doi.org/10.1103/PhysRevLett.77.3865}{Phys. Rev. Lett. \textbf{77}, 3865 (1996)}.

\bibitem{Liechtenstein1995} A. I. Liechtenstein, V. I. Anisimov, and J. Zaanen, Density-functional theory and strong interactions: Orbital ordering in Mott-Hubbard insulators, \href{https://doi.org/10.1103/PhysRevB.52.R5467}{Phys. Rev. B \textbf{52}, R5467 (1995)}.

\bibitem{Corso2014} A. Dal Corso, Pseudopotentials periodic table: From H to Pu, \href{https://doi.org/10.1016/j.commatsci.2014.07.043}{Comput. Mater. Sci. \textbf{95}, 337 (2014)}.

\bibitem{Marzhari2012} N. Marzari, A. A. Mostofi, J. R. Yates, I. Souza, and D. Vanderbilt, Maximally localized Wannier functions: Theory and applications, \href{https://doi.org/10.1103/RevModPhys.84.1419}{Rev. Mod. Phys. \textbf{84}, 1419 (2012)}.

\bibitem{Marzhari1997} N. Marzari and D. Vanderbilt, Maximally localized generalized Wannier functions for composite energy bands, \href{https://doi.org/10.1103/PhysRevB.56.12847}{Phys. Rev. B \textbf{56}, 12847 (1997)}.

\bibitem{Souza2001} I. Souza, N. Marzari, and D. Vanderbilt, Maximally localized Wannier functions for entangled energy bands, \href{https://doi.org/10.1103/PhysRevB.65.035109}{Phys. Rev. B \textbf{65}, 035109 (2001)}.

\bibitem{Mostofi2008} A. A. Mostofi, J. R. Yates, Y.-S. Lee, I. Souza, D. Vanderbilt, and N. Marzari, wannier90: A tool for obtaining maximally-localised Wannier functions, \href{https://doi.org/10.1016/j.cpc.2007.11.016}{Comput. Phys. Commun. \textbf{178}, 685 (2008)}.

\bibitem{Mostofi2014} A. A. Mostofi, J. R. Yates, G. Pizzi, Y.-S. Lee, I. Souza, D. Vanderbilt, and N. Marzari, An updated version of wannier90: A tool for obtaining maximally-localised Wannier functions, \href{https://doi.org/10.1016/j.cpc.2014.05.003}{Comput. Phys. Commun. \textbf{185}, 2309 (2014)}.

\bibitem{Pizzi2020} G. Pizzi, V. Vitale, R. Arita, S. Blügel, F. Freimuth, G. Géranton, M. Gibertini, D. Gresch, C. Johnson, T. Koretsune, J. Ibañez-Azpiroz, H. Lee, J.-M. Lihm, D. Marchand, A. Marrazzo, Y. Mokrousov, J. I. Mustafa, Y. Nohara, Y. Nomura, L. Paulatto, S. Poncé, T. Ponweiser, J. Qiao, F. Thöle, S. S. Tsirkin, M. Wierzbowska, N. Marzari, D. Vanderbilt, I. Souza, A. A. Mostofi, and J. R. Yates, Wannier90 as a community code: new features and applications, \href{https://doi.org/10.1088/1361-648X/ab51ff}{J. Phys. Condens. Mater. \textbf{32}, 165902 (2020)}.

\bibitem{Sancho1985} M. P. L. Sancho, J. M. L. Sancho, J. M. L. Sancho, and J. Rubio, Highly convergent schemes for the calculation of bulk and surface Green functions,  \href{https://doi.org/10.1088/0305-4608/15/4/009}{J. Phys. F: Met. Phys. \textbf{15}, 851 (1985)}.

\bibitem{Wu2018} Q. Wu, S. Zhang, H.-F. Song, M. Troyer, and A. A. Soluyanov, WannierTools: An open-source software package for novel topological materials,  \href{https://doi.org/10.1016/j.cpc.2017.09.033}{Comput. Phys. Commun. \textbf{224}, 405 (2018)}.

\bibitem{Lei2019} S. Lei, V. Duppel, J. M. Lippmann, J. Nuss, B. V. Lotsch, and L. M. Schoop, \href{https://doi.org/10.1002/qute.201900045}{Adv. Quantum Technol. \textbf{2}, 1900045 (2019)}.

\end{thebibliography}

\begin{thebibliography}{50}
\bibitem{Hohenberg1964S} P. Hohenberg and W. Kohn,  Inhomogeneous Electron Gas,  \href{https://doi.org/10.1103/PhysRev.136.B864}{Phys. Rev. \textbf{136,} B864 (1964)}.

\bibitem{Kohn1965S} W. Kohn and L. J. Sham, Self-Consistent Equations Including Exchange and Correlation Effects, \href{https://doi.org/10.1103/PhysRev.140.A1133}{Phys. Rev. \textbf{140}, A1133 (1965)}.

\bibitem{Blochl1994S} P. E. Blöchl, Projector augmented-wave method,  \href{https://doi.org/10.1103/PhysRevB.50.17953}{Phys. Rev. B \textbf{50,} 17953 (1994)}.

\bibitem{Kresse1994S} G. Kresse and J. Hafner, Ab initio molecular-dynamics simulation of the liquid-metal--amorphous-semiconductor transition in germanium, \href{https://doi.org/10.1103/PhysRevB.49.14251}{Phys. Rev. B \textbf{49}, 14251 (1994)}.

\bibitem{Kresse1996S} G. Kresse and J. Furthmüller, Efficient iterative schemes for ab initio total-energy calculations using a plane-wave basis set, \href{https://doi.org/10.1103/PhysRevB.54.11169}{Phys. Rev. B \textbf{54,} 11169 (1996)}.

\bibitem{Kresse1999S} G. Kresse and D. Joubert, From ultrasoft pseudopotentials to the projector augmented-wave method, \href{https://doi.org/10.1103/PhysRevB.59.1758}{Phys. Rev. B \textbf{59}, 1758 (1999)}.

\bibitem{Perdew1996S} J. P. Perdew, K. Burke, and M. Ernzerhof, Generalized Gradient Approximation Made Simple, \href{https://doi.org/10.1103/PhysRevLett.77.3865}{Phys. Rev. Lett. \textbf{77}, 3865 (1996)}.

\bibitem{Liechtenstein1995S} A. I. Liechtenstein, V. I. Anisimov, and J. Zaanen, Density-functional theory and strong interactions: Orbital ordering in Mott-Hubbard insulators, \href{https://doi.org/10.1103/PhysRevB.52.R5467}{Phys. Rev. B \textbf{52}, R5467 (1995)}.

\bibitem{Giannozzi2009S} P. Giannozzi, S. Baroni, N. Bonini, M. Calandra, R. Car, C. Cavazzoni, D. Ceresoli, G. L. Chiarotti, M. Cococcioni, I. Dabo, A. Dal Corso, S. de Gironcoli, S. Fabris, G. Fratesi, R. Gebauer, U. Gerstmann, C. Gougoussis, A. Kokalj, M. Lazzeri, L. Martin-Samos, N. Marzari, F. Mauri, R. Mazzarello, S. Paolini, A. Pasquarello, L. Paulatto, C. Sbraccia, S. Scandolo, G. Sclauzero, A. P. Seitsonen, A. Smogunov, P. Umari, and R. M. Wentzcovitch, QUANTUM ESPRESSO: a modular and open-source software project for quantum simulations of materials, \href{https://doi.org/10.1088/0953-8984/21/39/395502}{J. Phys. Condens. Mater. \textbf{21}, 395502 (2009)}.

\bibitem{Giannozzi2017S} P. Giannozzi, O. Andreussi, T. Brumme, O. Bunau, M. Buongiorno Nardelli, M. Calandra, R. Car, C. Cavazzoni, D. Ceresoli, M. Cococcioni, N. Colonna, I. Carnimeo, A. Dal Corso, S. de Gironcoli, P. Delugas, R. A. DiStasio, A. Ferretti, A. Floris, G. Fratesi, G. Fugallo, R. Gebauer, U. Gerstmann, F. Giustino, T. Gorni, J. Jia, M. Kawamura, H. Y. Ko, A. Kokalj, E. Küçükbenli, M. Lazzeri, M. Marsili, N. Marzari, F. Mauri, N. L. Nguyen, H. V. Nguyen, A. Otero-de-la-Roza, L. Paulatto, S. Poncé, D. Rocca, R. Sabatini, B. Santra, M. Schlipf, A. P. Seitsonen, A. Smogunov, I. Timrov, T. Thonhauser, P. Umari, N. Vast, X. Wu, and S. Baroni, Advanced capabilities for materials modelling with Quantum ESPRESSO, \href{https://doi.org/10.1088/1361-648X/aa8f79}{J. Phys. Condens. Mater. \textbf{29}, 465901 (2017)}.

\bibitem{Giannozzi2020S} P. Giannozzi, O. Baseggio, P. Bonfà, D. Brunato, R. Car, I. Carnimeo, C. Cavazzoni, S. de Gironcoli, P. Delugas, F. Ferrari Ruffino, A. Ferretti, N. Marzari, I. Timrov, A. Urru, and S. Baroni, Quantum ESPRESSO toward the exascale, \href{https://doi.org/10.1063/5.0005082}{J. Chem. Phys.  \textbf{152}, 154105 (2020)}.

\bibitem{Corso2014S} A. Dal Corso, Pseudopotentials periodic table: From H to Pu, \href{https://doi.org/10.1016/j.commatsci.2014.07.043}{Comput. Mater. Sci. \textbf{95}, 337 (2014)}.

\bibitem{Marzhari2012S} N. Marzari, A. A. Mostofi, J. R. Yates, I. Souza, and D. Vanderbilt, Maximally localized Wannier functions: Theory and applications, \href{https://doi.org/10.1103/RevModPhys.84.1419}{Rev. Mod. Phys. \textbf{84}, 1419 (2012)}.

\bibitem{Marzhari1997S} N. Marzari and D. Vanderbilt, Maximally localized generalized Wannier functions for composite energy bands, \href{https://doi.org/10.1103/PhysRevB.56.12847}{Phys. Rev. B \textbf{56}, 12847 (1997)}.

\bibitem{Souza2001S} I. Souza, N. Marzari, and D. Vanderbilt, Maximally localized Wannier functions for entangled energy bands, \href{https://doi.org/10.1103/PhysRevB.65.035109}{Phys. Rev. B \textbf{65}, 035109 (2001)}.

\bibitem{Mostofi2008S} A. A. Mostofi, J. R. Yates, Y.-S. Lee, I. Souza, D. Vanderbilt, and N. Marzari, wannier90: A tool for obtaining maximally-localised Wannier functions, \href{https://doi.org/10.1016/j.cpc.2007.11.016}{Comput. Phys. Commun. \textbf{178}, 685 (2008)}.

\bibitem{Mostofi2014S} A. A. Mostofi, J. R. Yates, G. Pizzi, Y.-S. Lee, I. Souza, D. Vanderbilt, and N. Marzari, An updated version of wannier90: A tool for obtaining maximally-localised Wannier functions, \href{https://doi.org/10.1016/j.cpc.2014.05.003}{Comput. Phys. Commun. \textbf{185}, 2309 (2014)}.

\bibitem{Pizzi2020S} G. Pizzi, V. Vitale, R. Arita, S. Blügel, F. Freimuth, G. Géranton, M. Gibertini, D. Gresch, C. Johnson, T. Koretsune, J. Ibañez-Azpiroz, H. Lee, J.-M. Lihm, D. Marchand, A. Marrazzo, Y. Mokrousov, J. I. Mustafa, Y. Nohara, Y. Nomura, L. Paulatto, S. Poncé, T. Ponweiser, J. Qiao, F. Thöle, S. S. Tsirkin, M. Wierzbowska, N. Marzari, D. Vanderbilt, I. Souza, A. A. Mostofi, and J. R. Yates, Wannier90 as a community code: new features and applications, \href{https://doi.org/10.1088/1361-648X/ab51ff}{J. Phys. Condens. Mater. \textbf{32}, 165902 (2020)}.

\bibitem{Wu2018S} Q. Wu, S. Zhang, H.-F. Song, M. Troyer, and A. A. Soluyanov, WannierTools: An open-source software package for novel topological materials,  \href{https://doi.org/10.1016/j.cpc.2017.09.033}{Comput. Phys. Commun. \textbf{224}, 405 (2018)}.

\bibitem{Sancho1985S} M. P. L. Sancho, J. M. L. Sancho, J. M. L. Sancho, and J. Rubio, Highly convergent schemes for the calculation of bulk and surface Green functions,  \href{https://doi.org/10.1088/0305-4608/15/4/009}{J. Phys. F: Met. Phys. \textbf{15}, 851 (1985)}.



\end{thebibliography}
\end{document}